\documentclass[
 reprint,
 amsmath,amssymb,
 aps,
]{revtex4-2}

\usepackage{graphicx}
\usepackage{dcolumn}
\usepackage{bm}

\begin{document}

\preprint{APS/123-QED}

\title{Highly-Excited Rydberg Excitons in Synthetic Thin-Film Cuprous Oxide}

\author{Jacob DeLange}
 \email{jdelange@purdue.edu}
 \affiliation{%
 Department of Physics, Purdue University, West Lafayette IN 47907, USA
}%
\author{Kinjol Barua}%
 \affiliation{%
 Elmore Family School of Electrical and Computer Engineering, Purdue University, West Lafayette IN 47907, USA
}%
\author{Val Zwiller}
\author{Stephan Steinhauer}%
 \affiliation{Department of Applied Physics, KTH Royal Institute of Technology, SE 106 91 Stockholm, Sweden}
\author{Hadiseh Alaeian}%
 \affiliation{%
 Elmore Family School of Electrical and Computer Engineering, Purdue University, West Lafayette IN 47907, USA
}%
 \affiliation{%
 Department of Physics, Purdue University, West Lafayette IN 47907, USA
}%

\date{\today}

\begin{abstract}
Cuprous oxide (Cu${}_2$O) has recently been proposed as a promising solid-state host for excitonic Rydberg states with large principal quantum numbers ($n$), whose exaggerated wavefunction sizes ($\propto n^2)$ facilitate gigantic dipole-dipole ($\propto n^4$) and van der Waals ($\propto n^{11}$) interactions, making them an ideal basis for solid-state quantum technology. Synthetic, thin-film Cu${}_2$O samples are of particular interest because they can be made defect-free via carefully controlled fabrication and are, in principle, suitable for the observation of extreme single-photon nonlinearities caused by the Rydberg blockade. Here, we present spectroscopic absorption and photoluminescence studies of Rydberg excitons in synthetic Cu${}_2$O grown on a transparent substrate, reporting yellow exciton series up to $n = 7$. We perform these studies at powers up to 2 mW and temperatures up to 150 K, the highest temperature where Rydberg series can be observed. These results open a new portal to scalable and integrable on-chip Rydberg-based quantum devices.
\end{abstract}

\maketitle


\section{Introduction}~\label{Introduction}

Photons are one of the most promising candidates for exploiting quantum mechanical laws in the era of the second quantum revolution.
Compared to particles such as atoms or ions, they are less susceptible to environmental perturbations and can be efficiently generated, measured, controlled, and transmitted over long distances. However, the lack of strong interactions between photons significantly hinders their application for the development of large quantum networks and scalable quantum technologies~\cite{Wang2020, Vigliar2021, Bourassa2021}. Excitons, the solid-state equivalent of the hydrogen atom, can mediate interactions between photons and create giant optical nonlinearities \cite{Bayer2020}. Besides, because excitons exist in solid-state platforms, which are inherently robust, miniature, and scalable, they pave the way for novel quantum devices. Excitons are composed of a positive hole and a negative electron, resulting in a hydrogen-like energy level structure governed by the Rydberg formula~\cite{Rydberg1890}

\begin{equation}
    E_n = E_g - \frac{Ry}{(n - \delta_n)^2} \, ,
\label{RydbergFormula}
\end{equation}

\noindent
where $E_g$ is the bandgap energy, $Ry$ is the binding energy, $n$ is the principal quantum number of the state, and $\delta_n$ is the quantum defect, which describes perturbations caused by the screening of Coulombic interactions within Cu${}_2$O's lattice and the non-parabolicity of Cu${}_2$O's band structure \cite{Bayer2020}. States with high principal quantum numbers, known as Rydberg states, are linked to long lifetimes which are proportional to $n^3$, meaning that highly excited states could be used to obtain long coherence times on the order of nanoseconds~\cite{Bayer2020, Orfanakis2021}. Moreover, Rydberg excitons have wavefunction sizes which scale as $n^2$, allowing them to interact via long-range dipole-dipole and Van der Waals interactions, which scale as $n^4$ and $n^{11}$, respectively~\cite{Gallagher1994, Sibalic2018}. These interaction can be large enough to perturb the energy level of nearby atoms or excitons to the point that they no longer have the same excitation frequency as the single isolated atom or exciton~\cite{Urban2009}. This phenomenon, known as \textit{Rydberg blockade}, has been the backbone of several atomic Rydberg quantum systems such as all-optical transistors~\cite{Gorniaczyk2014, Tiarks2014}, high fidelity two-qubit gates~\cite{Levine2018, Saffman2016}, and analog quantum simulators~\cite{Scholl2021, Ebadi2021}.

\begin{figure*}[htpb!]
\centering\includegraphics[width=14cm]{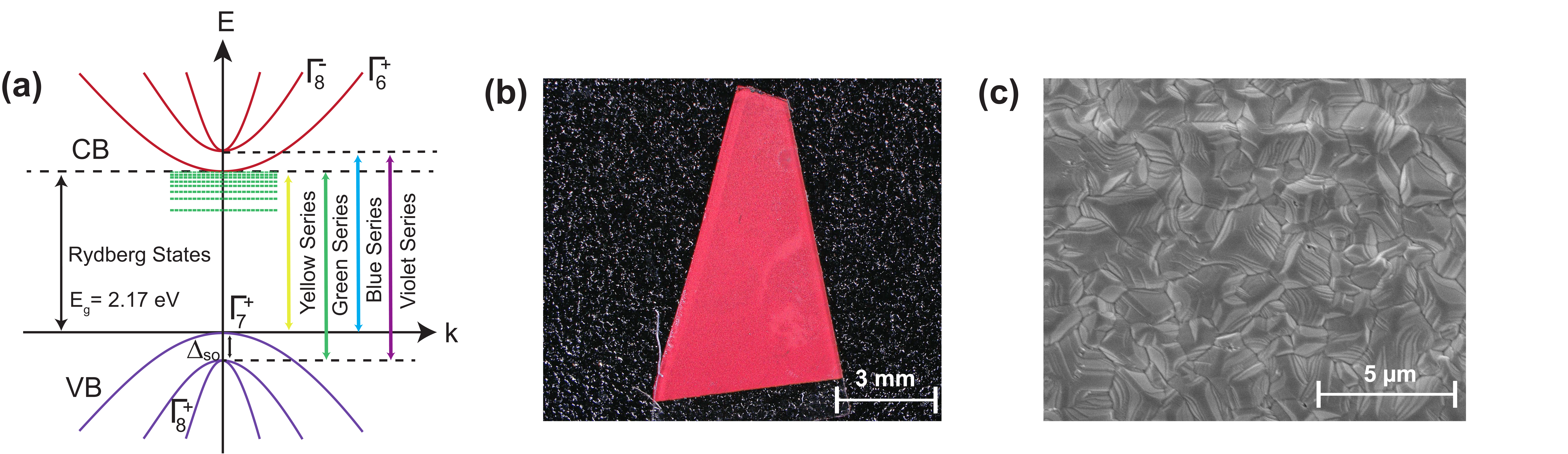}
\caption{Overview of synthetic Cu${}_2$O sample. (a) Band structure of Cu${}_2$O showing green, yellow, blue, and violet Rydberg series. (b) Microscope image of the synthetic sample on a transparent substrate. (c) High-resolution scanning electron microscope image of the sample showing Cu$_2$O micro-crystals.}
\label{Figure1_final}
\end{figure*}

Here, we are interested in the semiconductor cuprous oxide, which has been recently emerged as an ideal host for Rydberg excitons~\cite{Bayer2020}, with properties necessary to bring the strength of blockade interactions to a solid-state platform~\cite{Walther2018, Zhao2021, Su2021}. As shown in Fig.~\ref{Figure1_final}(a), it is a direct bandgap semiconductor ($E_g \approx 2.17$ eV) with a symmetric, cubical lattice which suppresses coupling to phonon modes and a large binding energy (around 98 meV) which allows it to host a large number of Rydberg states without them undergoing the thermal ionization. Its two topmost valence bands (VBs) ($\Gamma_8^+$  and $\Gamma_7^+$) are created from Cu 3d electrons, whereas the lowest and highest conduction bands (CBs) ($\Gamma_6^+$ and $\Gamma_8^-$) result from 4s electrons in Cu and 2p electrons in O, respectively~\cite{Elliott1961}. Transitions between the bands lead to four exciton series named the yellow, green, blue, and violet series for the colors of their emitted photons. Here, we are interested in the yellow series, whose valence band has the same parity as its conduction band, meaning that while p-excitons are dipole-allowed to directly recombine to the valence band, the 1s-excitons are dipole-forbidden. When considering the fine structure, the 1s-orthoexciton recombination is only quadrupole allowed while the 1s-paraexciton recombination is forbidden up to all orders~\cite{Otter2007,Bayer2020}. This property sets Cu${}_2$O apart even from other Rydberg-compatible materials, such as transition metal dichalcogenides (TMDCs) that sport giant binding energies on the order of hundreds of meV~\cite{Zhu2015, Park2018}, but have ground state lifetimes on the order of picoseconds due to their dipole-allowed states~\cite{Robert2016}. 

So far, a preponderance of the studies on Rydberg excitons in Cu${}_2$O have been focused on bulk samples obtained from the Tsumeb mine in Namibia and mechanically polished to ensure surface smoothness~\cite{Kazimierczuk2014, Takahata2018, Mund2018, Kang2021, Versteegh2021, Gallagher2022, Orfanakis2022}. In these studies, states up to $n = 30$ have been reported~\cite{Kazimierczuk2014}. However, a growing number of studies have been conducted on synthetic samples, where states up to $n = 10$ have been observed~\cite{Steinhauer2020, Lynch2021}. These studies not only promise to realize Cu$_2$O's potential for scalability, but also to allow the coupling of excitons to on-chip nanophotonic circuits. Further, in thin-film samples whose thicknesses are smaller than the blockade radius, only one Rydberg exciton can be excited, leading to an extreme nonlinearity at the single-photon level. This nonlinearity could not only be harnessed in quantum devices such as single-photon sources or low-intensity optical switches, but could also applied to the general development of efficient, nonlinear quantum optical devices~\cite{Khazali2017, Walther2018, Walther2022}. In this article, we present the first spectroscopic characterization of Rydberg excitons on a synthetic thin-film sample of Cu$_2$O on a fused silica substrate, as well as the effects of temperature and excitation power on these resonances.

\section{Results and Discussions}
\subsection{Sample Preparation}~\label{Sample Preparation}

To synthesize the sample, a 700 nm Cu film was deposited on the substrate via e-beam evaporation, with a 5 nm Ti layer in between the Cu and the substrate to increase adhesion. Next, the Cu was oxidized at 850 ${}^\circ$C and 1 mbar pressure to get Cu${}_2$O~\cite{Steinhauer2020}. Fig.~\ref{Figure1_final}(b) shows a microscope image of the finished sample. To characterize the sample, we examined it with a scanning electron microscope (SEM), as shown in Fig.~\ref{Figure1_final}(c). As can be seen, the Cu$_2$O forms a microcrystalline thin-film on the substrate.

\begin{figure*}[htpb!]
\centering\includegraphics[width=14cm]{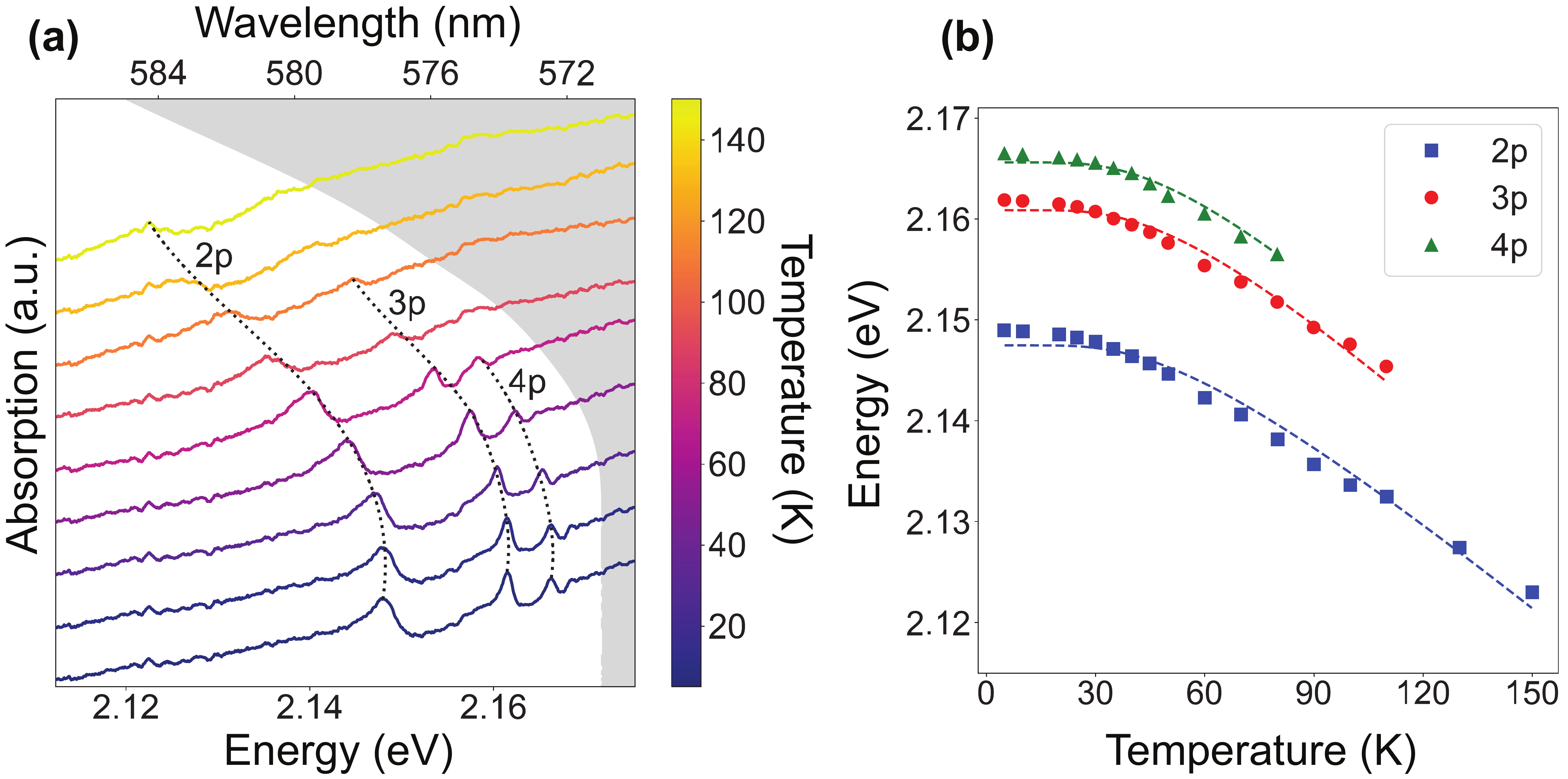}
\caption{Results from optical density measurement. (a) Temperature-dependent OD of the synthetic Cu$_2$O film. Line colors correspond to different temperatures. The grayed out region represents energies above the bandgap energy, determined using Elliott's model, see Eq. (\ref{ElliotModel}). (b) Rydberg exciton 2p, 3p, and 4p peaks as a function of temperature. The dashed lines represent the least squares fit from Elliott's model.}
\label{ODSummaryFig}
\end{figure*}


\subsection{Optical Density Measurement}~\label{OpticalDensityMeasurement}

The optical density (OD) measurement was performed by passing light from a broadband white light source (Thorlabs SLS201L) through the sample. The light was focused onto the sample with an objective lens with NA = 0.42 (Mitutotyo Plan Apo 20x). The transmitted light was collected with an objective lens of the same model and sent to a spectrometer (Princeton Instruments HRS-750, 1200 gratings/mm, 50 $\mu$m slit width, resolution of 0.036 nm). The measurement was repeated at temperatures from 5-150 K with a precision of 0.25 K. Fig.~\ref{ODSummaryFig}(a) shows the results from this measurement for the yellow exciton series. At low temperatures, resonances up to 4p were observed. As the temperature was increased, the center wavelengths of the exciton resonances were red shifted and broadened until they could no longer be resolved. At temperatures higher than 150 K, no distinct exciton resonances could be observed. The shifts in the center wavelengths of the peaks can be attributed to changes in the bandgap energy, binding energy, and quantum defects of Cu$_2$O as a function of temperature. The change in the bandgap energy, $E_g$, arises from the thermal expansion of the crystal lattice and phonon-electron interactions~\cite{Elliott1957}. The binding energy, $Ry$, being proportional to the reduced mass of the exciton, is also temperature-dependent due to changes in the electronic bands curvature~\cite{Bayer2020}. Finally, the quantum defect, which arises in part due to the non-parabolicity of the bands, will likewise vary with temperature as the band structures are modified~\cite{Schone2017}.

The changes in the first two parameters can be summarized by Elliott's model (\ref{ElliotModel}), which assumes that shifts are dominated by continuous absorption from $\Gamma_3^-$ phonons~\cite{Elliott1957},

\begin{equation}
\begin{aligned}
E_g(T) = E_{g0} + E_{gT} \Big[\coth{\Big(\frac{\hbar \omega_3}{2 k_B T}\Big)} - 1\Big] \, ,\\ 
Ry(T) = Ry_{0} + Ry_{T} \Big[\coth{\Big(\frac{\hbar \omega_3}{2 k_B T}\Big)} - 1\Big] \, , 
\label{ElliotModel}
\end{aligned}
\end{equation}

\noindent
with $\ \hbar\omega_3 = 13.6$ meV.

$E_{g0}$ and $Ry_0$ are temperature-independent terms that represent the low-temperature limit of the bandgap and binding energies, respectively and $E_{gT}$ and $Ry_T$ capture the temperature effects. To fit this model to our data, the Rydberg energies were extracted for each temperature using least squares fitting to an asymmetric Fano lineshape~\cite{Fano1961}

\begin{equation}
    \alpha_n(E) = C_n \frac{\frac{\Gamma_n}{2} + 2 q_n (E - E_n)}{\big(\frac{\Gamma_n}{2}\big)^2 + (E - E_n)^2}\, ,
\end{equation}

\noindent
where $E_n$ is the center energy of the n$^{th}$ excitonic level, $\Gamma_n$ is the corresponding linewidth, $C_n$ is proportional to the oscillator strength, and $q_n$ is an asymmetry factor modeling the interference between narrow optical transitions and the phonon continuum~\cite{Toyozawa1964}. 

As is visible in Fig.~\ref{ODSummaryFig}(a), there is a background from continuum absorption which increases at higher energies. This background was fitted with an Urbach tail, which signifies an increase in absorption near the bandgap energy caused by an increasing density of states in that energy range. This increase is described by an exponential function,

\begin{equation}
    \alpha_{U}(E) = \alpha_0 \exp{\Big(\frac{E - E_g}{E_u}\Big)}\, ,
\end{equation}

\noindent
where $E_g$ is the bandgap energy, $\alpha_0$ is the magnitude of the continuum absorption, and $E_u$ is the Urbach energy~\cite{Urbach1953, Cody1981}.

Elliott's model was used to fit the center energy as a function of temperature, taking into account the 2p, 3p, and 4p peaks simultaneously. This fit is shown in Fig.~\ref{ODSummaryFig}(b). For simplicity, we ignored the quantum defects for all three resonances at all temperatures. This is just an approximation since the quantum defects vary with $n$ and are expected to be temperature-dependent as well. Typical methods to extract the quantum defect as a function of $n$, such as those used in~\cite{Kang2021}, require the observation of high energy peaks whose quantum defects approach a constant value (see supplementary materials for more details). Theoretical calculations show that this trend does not emerge until $n \gtrapprox 10$~\cite{Schone2017}, but here we were only able to observe peaks up to $n = 4$, making the methods inapplicable here. However, in~\cite{Kang2021} the authors note that while assuming $\delta_n(T) = 0$ is simplistic, it yields results which agree with both the literature and more detailed analyses which includes the quantum defect corrections. As such, we have chosen to employ this approximation here as well. From this fit, we have extracted the parameters shown in Table~\ref{ODElliottFitParams}.

\begin{table}
    \caption{Fit Parameters from Elliott Model}
    \label{ODElliottFitParams}
    \centering
        \begin{tabular}{|c c c c|} 
        \hline
        $E_{g0}$ (meV) & $E_{gT}$ (meV) & $Ry_0$ (meV) & $Ry_T$ (meV) \\ [0.5ex] 
        \hline\hline
        2171.7 $\pm$  0.04 & -29.5 $\pm$ 1.75 & 96.8 $\pm$ 2.12 & -20.9 $\pm$ 8.09\\ 
        \hline
        \end{tabular}
\end{table}

The values of $E_{g0}$, $E_{gT}$, and $Ry_0$ are in agreement with the literature, but the extracted $Ry_T$ is different~\cite{Kazimierczuk2014, Kang2021}. It must be noted that ignoring $\delta_n$ is valid for describing the temperature dependence of the bandgap energy, but breaks down for the binding energy, so we attribute this particular discrepancy to the errors caused by this assumption. Overall, the accurate observations of $E_{g0}$, $E_{gT}$, and $Ry_0$ confirm that the observed absorption lines indeed arise from Rydberg excitons in the thin-film Cu$_2$O.

\subsection{Photoluminescence Measurement}~\label{PhotoluminescenceMeasurement}

Non-resonant photoluminescence (PL) measurements were performed with a 532 nm laser to excite electrons above the bandgap and create bound electron-hole pairs which form excitons as they relax to lower energies. The laser was focused down to a spot size of 3.1 $\mu$m full width at half maximum using an objective lens (Mitutoyo Plan Apo 20x). The reflected PL from the excitons was collected with the same objective and sent to the spectrometer. A longpass filter (Semrock LP03-532RE-25) and a dichroic mirror (Thorlabs DMLP567) were used to block the reflected 532 nm laser light from reaching the spectrometer. Spectra were taken for a variety of sample temperatures and incident laser powers. Temperature-dependent data was taken at an incident laser power of 50 $\mu$W at temperatures from 5-150 K. Power-dependent data was taken with the sample held at a constant temperature of 5 K at laser powers ranging from 50 $\mu$W to 2 mW.

\begin{figure*}[htpb!]
\centering\includegraphics[width=13cm]{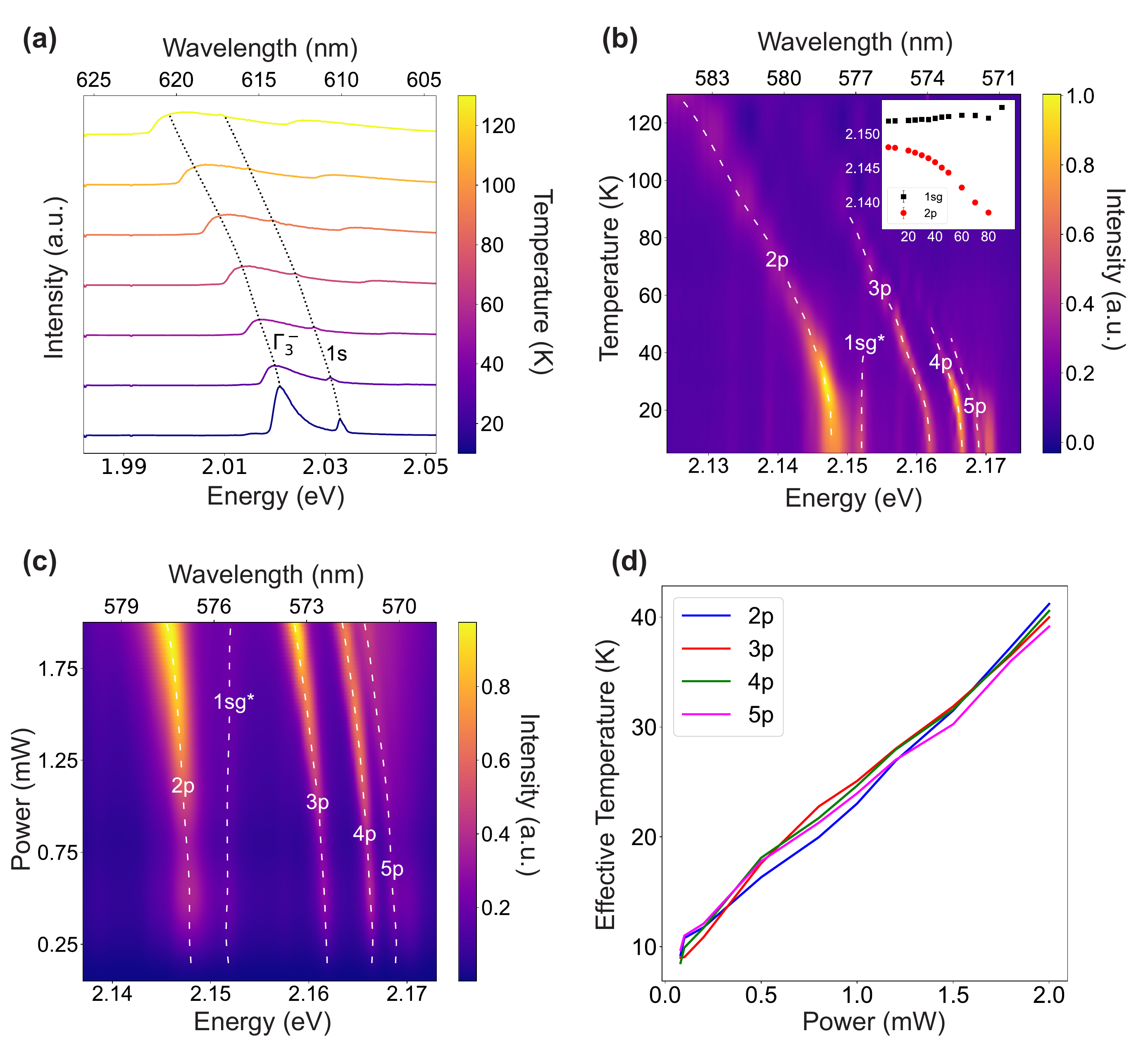}
\caption{Summary of results from PL measurement. (a) Temperature varying PL from phonon replica region. (b) Temperature varying PL from yellow exciton series. The line labeled as ``1s${}_{\textrm{g}}^{*}$" has often been attributed to the green 1s resonance, but as is shown in the inset, its center energy does not red shift with increasing temperature, raising doubts about this label. (c) Power dependent PL spectrum from synthetic Cu${}_2$O sample. (d) Interpolation of effective temperature induced by heating from 532 nm laser.}
\label{PLSummaryFig}
\end{figure*}

Fig.~\ref{PLSummaryFig}(a) shows the results from the phonon replica region of the spectrum, where the quadrupole-allowed and $\Gamma_3^-$ phonon-assisted relaxations of the yellow 1s-orthoexciton state can be observed as Fano and Boltzmann-tailed peaks, respectively. At higher temperatures one can see an anti-stokes phonon-assisted transition appearing at higher energies. As indicated by the black dotted line, these peaks red shift with temperature in a similar manner to the yellow excitons obtained from the OD measurement. Besides these transitions, the information from this spectral range is useful for gauging the prevalence of metallic impurities in the sample. In~\cite{Jang2006}, it is demonstrated that excitons bound to these impurities flouresce at energies between 1.99 and 2.01 eV, with intensities on the same order of magnitude as the $\Gamma_3^-$ phonon-assisted transition. In Fig.~\ref{PLSummaryFig}(a), the fact that no such features can be observed, even at low temperatures, demonstrates that the synthetic film is impurity-free.

The PL from the phonon replica peaks is also several orders of magnitude brighter than the PL from the yellow exciton series~\cite{Takahata2018}, so we used it to find the most optically active piece of our sample. Using a nano-positioner we scanned the sample through an 1800 $\times$ 2000 $\mu$m region in increments of $400$ $\mu$m, recording the brightness of the $\Gamma_3^-$ and 1s-orthoexciton peaks at each position. Once the brightest spot was determined via this coarse analysis (with the brightest spot being roughly twice as bright as the dimmest), a more fine analysis was conducted by scanning over a 10 $\times$ 10 $\mu$m region in increments of 1 $\mu$m. Whereas the coarse analysis yielded substantial variation in the brightness of these peaks, the fine analysis yielded no more than 5\% change.

At the brightest point on the sample at a temperature of 5 K, Rydberg states up to n = 7p were clearly distinguishable. The peaks were fit with Fano line shapes using a least squares algorithm, though for PL, an Urbach tail background was not required. Figure~\ref{PLSummaryFig}(b) shows the results from the temperature-dependent study. As in the OD measurements, the center wavelengths of the yellow exciton peaks observed in PL red shift as temperature increases. A similar trend can be observed in the power-dependent data shown in Fig.~\ref{PLSummaryFig}(c). The change in the resonance wavelength can be attributed to two possible effects: temperature change due to laser absorption and exciton-exciton interactions caused by the high exciton density created at high laser powers. As already discussed in Sec.~\ref{OpticalDensityMeasurement}, the former effect manifests as a red shift, while the latter is expected to manifest as a blue shift since it is caused by repulsive Van der Waals interactions~\cite{Sibalic2018}.

To differentiate between these two effects, we performed an interpolation, using the center energies from the power-dependent measurement to extract an effective temperature as a function of the laser power. Separate interpolations were performed for the 2p, 3p, 4p, and 5p peaks. Higher energy peaks up to at least n = 7p were also visible, but due to their overlap with each other, their resonance wavelength could not be reliably extracted. As can be seen in Fig.~\ref{PLSummaryFig}(d), all four interpolations follow the same trend. If there were a significant contribution from exciton-exciton interactions, the interpolations from higher energy peaks would be expected to deviate from the lower energy levels, as excitons in higher energy states interact more strongly due to the overlap of their extended wavefunctions. Combined with the fact that the interpolations are linear, this indicates that the energy shifts due to temperature change dominated over those from the exciton-exciton interaction, which is not surprising due to the strong absorption of photons above the bandgap.

In both Fig.~\ref{PLSummaryFig}(b) and \ref{PLSummaryFig}(c), another peak between the 2p and 3p resonances (at E = 2.152 eV for T = 5 K) labeled as 1s$_{\textrm{g}}^{*}$) can be observed which is not a part of the yellow exciton series. This resonance can be attributed to the 1s state of the green excitons as its low-temperature energy matches with theoretical calculations of the green series~\cite{Mund2018, Schweiner2017, Rommel2021}. The origin of this resonance and its behavior are currently under investigation and will be reported elsewhere.

\begin{figure*}[htpb!]
\centering\includegraphics[width=13cm]{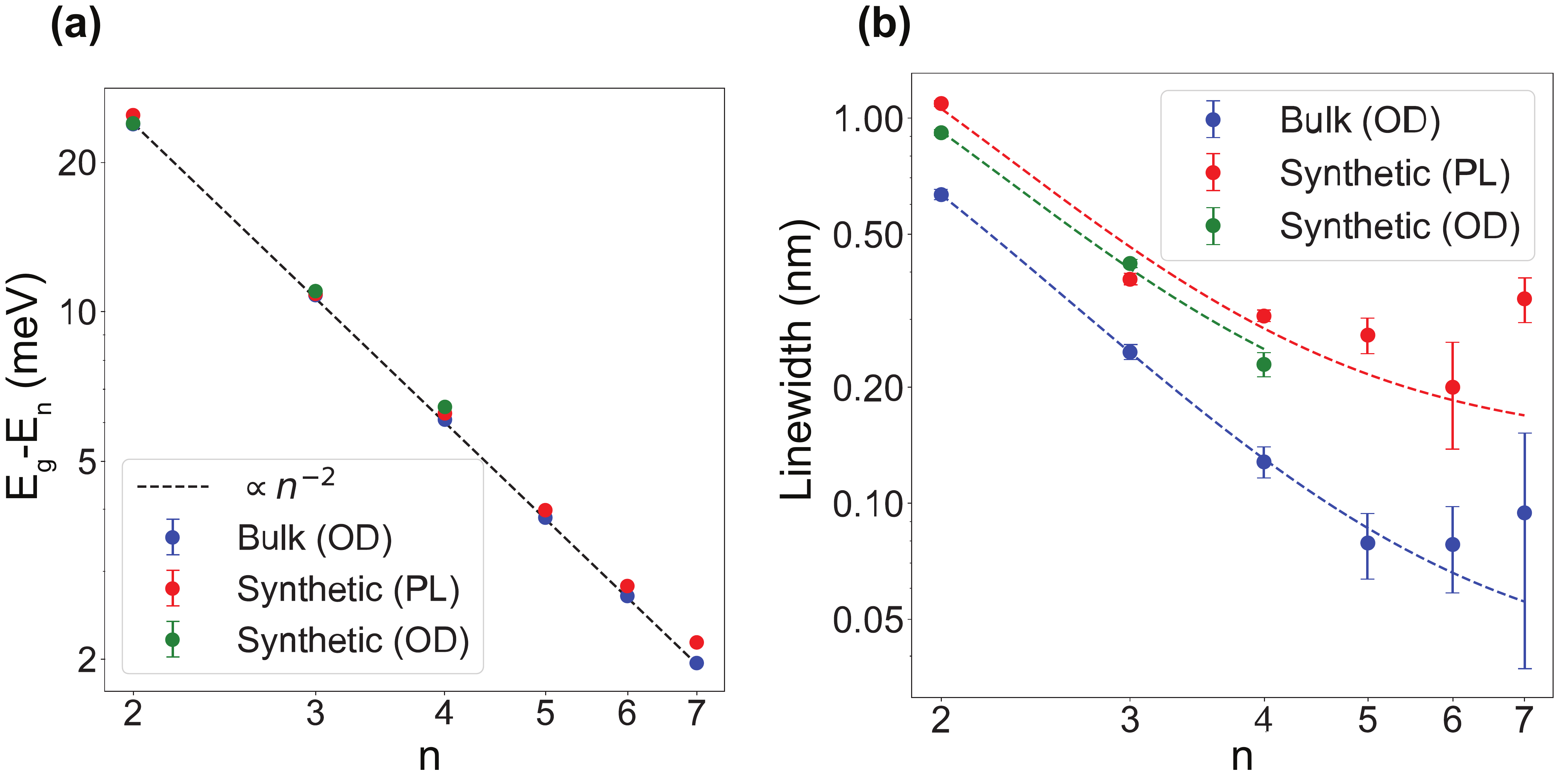}
\caption{Trends of various resonance parameters vs. principal quantum number. (a) Center energy as a function of $n$, with an $n^{-2}$ trendline overlaid. (b) Linewidth as a function of $n$. The dashed lines represent fits of the model given in Eq.~(\ref{LinewidthModel}) for the data series of the same color.}
\label{TrendlineFig}
\end{figure*}

In Fig.~\ref{TrendlineFig}, we compare the properties of the synthetic sample with those of natural Cu${}_2$O. The Rydberg exciton resonance energies and linewidths are plotted as a function of $n$ for data obtained from PL and OD measurements of the synthetic sample at the lowest temperature of 5 K and the lowest laser power of 50 $\mu$W, as well as for data from OD measurements of a natural bulk sample (see supplementary materials). 

As noted in Eq. (\ref{RydbergFormula}), resonance energies approach the bandgap energy as $n^{-2}$ (black dashed line). Deviation from this trend can be caused by confinement effects, whereby the wavefunction of a high energy exciton is comparable to the Cu$_2$O film thickness~\cite{Konzelmann2019}. The p-series exciton wavefunction size can be estimated as~\cite{Orfanakis2021}

\begin{equation}
    r_n = a_b (3n^2 - 2)\, ,
\end{equation}

\noindent
where $a_b$ is the Bohr radius of the yellow excitons with a value of 1.11 nm~\cite{Kavoulakis1997}. Thus, even for the 7p state, the highest state observed in this work, the wavefunction is only about 160 nm large and thus would be too small to experience noticeable perturbations from the confinement. This is verified in Fig.~\ref{TrendlineFig}(a), where all data points closely follow the $n^{-2}$ trendline.

As for the linewidths, in ideal Rydberg states, e.g. in atoms, they scale as $n^{-3}$, but as can be observed in Fig.~\ref{TrendlineFig}(b) they seem to reach to a plateau in a matter which agrees with previous results~\cite{Kazimierczuk2014}. The $n$-dependent behavior can be modeled as

\begin{equation}
    \Gamma(n) = \alpha\frac{n^2 - 1}{n^5} + \beta\, ,
\label{LinewidthModel}
\end{equation}

\noindent
where $\alpha$ is a proportionality constant and $\beta$ represents a minimum value which the linewidths approach as $n$ goes to infinity~\cite{Kruger2020}. This model was used to fit the data from each series shown in Fig.~\ref{TrendlineFig}(b). The high-$n$ asymptotes ($\beta$) for each series are reported in Table~\ref{BroadeningParams}. Similar linewidth saturation behavior has been observed for Rydberg atoms interacting with dense background gases~\cite{Schlagmuller2016}, a phenomenon attributed to perturbations of Rydberg electrons via the background gas~\cite{Liebisch2016}. In PL measurements, excitons in Cu${}_2$O interact with a background of electron-hole plasma as well as a high density of 1s excitons, so by analogy with the atomic Rydberg, this large-$n$ limit may arise from exciton-plasma interactions as well as the collision of high-energy and ground state excitons~\cite{Walther2020, Kitamura2017}.

\begin{table}
    \caption{Broadening linewidths, given by $\beta$ in Eq. (\ref{LinewidthModel}), for various measurements of Rydberg excitons in Cu${}_2$O (nm)}
    \label{BroadeningParams}
    \centering
        \begin{tabular}{|c c c|} 
        \hline
        Bulk (OD) & Synthetic (PL) & Synthetic (OD)\\ [0.5ex] 
        \hline\hline
        0.061 $\pm$ 0.009 & 0.181 $\pm$ 0.029 & 0.181 $\pm$ 0.043\\
        \hline
        \end{tabular}
\end{table}

\section{Conclusion}~\label{conclusion}

We have presented the first optical spectroscopy measurements of Rydberg excitons in a synthetic, thin-film sample of Cu$_2$O grown on a transparent fused silica substrate. In the yellow exciton region of the spectrum, we have reported Rydberg peaks up to n = 4p and n = 7p for OD and PL measurements, respectively, and shown that their resonance energies scale as $n^{-2}$. Additionally, we have studied their temperature-dependent behavior, showing that it is well-explained by Elliott's model, as has been previously reported for natural bulk Cu$_2$O crystals~\cite{Kang2021}. In a similar vein, we have shown that spectral variations induced by changes in excitation power are described properly by heating induced via optical absorption.

This study lays the groundwork for understanding the interaction of Rydberg excitons with their environment, such as the variation of quantum defects with temperature and the broadening induced by exciton interactions with electron-hole plasma. The adverse heating effects obscuring the exciton-exciton interactions can be mitigated by using a pulsed laser which generates high exciton densities while maintaining a low lattice temperature, allowing the signature of exciton interactions to be observed. While the blockade effect has recently been observed in bulk Cu${}_2$O samples~\cite{Heckotter2021}, achieving that limit in a synthetic  micro-crystal would be a key step towards complex solid-state Rydberg quantum systems.

Semiconductor Rydberg physics is still an emerging field, and our demonstration of Rydberg excitons in synthetic Cu${}_2$O opens the door for their study in more complex settings such as nanophotonic circuits which imprint large optical nonlinearities, or optical cavities which facilitate strongly-interacting Rydberg exciton-polaritons~\cite{Orfanakis2022}. In the long-term, this would facilitate the development of scalable, on-chip devices such as photonic quantum gates and on-demand single-photon sources.

\begin{acknowledgments}
The authors acknowledge stimulating discussions with H. Ohadi, S. Scheel, and R. L\"ow.
\end{acknowledgments}

\section*{Disclosures}
The authors declare no conflicts of interest.

\section*{Data Availability Statement}
Data can be obtained from the corresponding author upon reasonable request.

\appendix

\section{Bayesian Reconstruction of Optical Density Measurement on Bulk Cu${}_2$O}

In the excitonic spectra of Cu${}_2$O, high energy peaks become hard to distinguish because their oscillator strengths decreases as $n^{-3}$ and their spacing decreases as $n^{-2}$, causing the peaks to overlap as they approach the bandgap~\cite{Bayer2020}. In these cases, least squares fitting algorithms alone are not sufficient for discerning different exciton resonances. This can be done more effectively using Bayesian reconstruction~\cite{Tokuda2016SimultaneousEO}  which helps to identify even those features which have intensities below the noise level.

In Bayesian reconstruction, a model with fit parameters $\theta$ can be fitted to a dataset $D$ by leveraging Bayes' law as

\begin{equation}
    P(\theta|D)\propto P(D|\theta)P(\theta) \, ,
\end{equation}

\noindent
where $P(\theta|D)$ represents the probability of a given set of fit parameters given the measured dataset, $P(\theta)$ is the prior probability of the fit parameters, and $P(D|\theta)$, the probability of a dataset being measured, given by the equation

\begin{equation}
    P(D|\theta) \propto \exp[-\frac{n\epsilon(\theta)}{\sigma^2}]\, ,
\end{equation}

\noindent
where $\sigma$ is the standard deviation of the background noise, assumed to be Gaussian, and $\epsilon(\theta)$ is the mean-squared error (MSE) between the dataset and the model. For example \emph{Metropolis algorithm}, which randomly samples from complex distributions can be used to determine the underlying distribution of the fit parameters~\cite{Iwamitsu2016} . However, this method requires the assumption of an underlying noise level which is not previously known and does not provide a metric for distinguishing between different types of models. To resolve this issue, the \textit{Replica Exhcange Monte Carlo (RXMC) method} is used to test multiple potential noise levels simultaneously and compare the likelihoods of different models \cite{okada2020}.

\subsection{Spectral Decomposition of Bulk Cu${}_2$O using Bayesian Estimation}

\begin{figure*}[htpb!]
\centering\includegraphics[width=13cm]{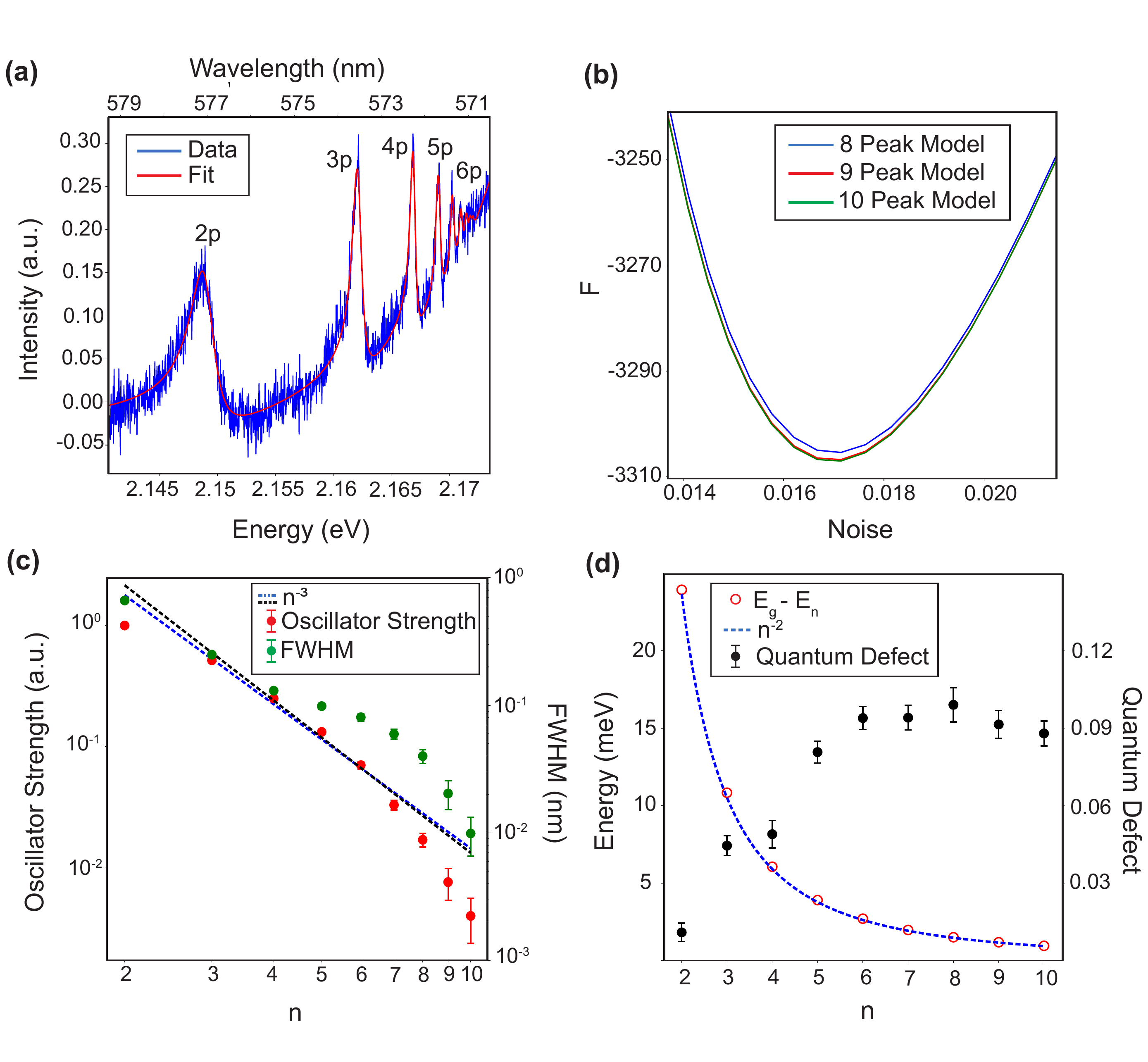}
\caption{Overview of Bayesian reconstruction from bulk Cu${}_2$O sample. (a) Raw and fitted absorption spectrum showing 2-6p excitonic resonances. (b)  $F$ vs. noise level for 8-10 peak models. (c) Oscillator strengths and spectral full widths at half maximum (FWHM) compared to a $n^{-3}$ trendline. (d) Rydberg exciton resonance energies and quantum defects from n = 10 resonance peaks model. The blue dotted curve is a trendline showing $n^{-2}$ dependence.}
\label{BayesianFig}
\end{figure*}

The spectroscopic absorption data set $D$ was measured from a bulk natural Cu${}_2$O crystal (sample was provided by Ohadi's group at the University of St Andrews, UK) sandwiched between two CaF${}_2$ windows. Figure~\ref{BayesianFig}(a) shows the raw and fitted absorption spectra where n = 2-6 p-exciton peaks are clearly distinguishable. More peaks are visible, but their spectral overlap makes it difficult to investigate them with certainty.

To resolve this issue, Bayesian reconstruction was used. Multiple models (corresponding to different numbers of peaks) were fit to the same dataset using the RXMC method. Each peak was fit with a Fano function, and the exponential behaviour at the tail of the spectrum was modeled by an Urbach tail. The output of the RXMC method is a parameter $F$, which is found by averaging the probability of each set of fit parameters over the entire distribution found during the execution of the algorithm. $F$ is analogous to the Helmholtz free energy, which is minimized by the most probable configuration in statistical mechanics. Similarly, the $F$ found by the RXMC method is minimized by the most probable set of fit parameters. A unique value of $F$ is given for each background noise tested, yielding the curves shown in Fig.~\ref{BayesianFig}(b). Each model generates a curve, and the model with the lowest minimum $F$ is deemed the most probable. Here, the ten peak model gives the lowest value for $F$, which means there are 10 distinct resonance peaks in the absorption spectrum shown in Fig.~\ref{BayesianFig}(a). Bayesian reconstruction allows us to determine the presence of these peaks despite the fact that they cannot be distinguished visibly because of their intensity falling below the noise floor.

To test the validity of these results, we plotted the values of the oscillator strengths, linewidths, resonance energies, and quantum defects as functions of $n$. Figure~\ref{BayesianFig}(c) shows the oscillator strengths and spectral linewidths of the 10 resonance peak model. The oscillator strengths follow $n^{-3}$ tendency as shown by the dotted line. However, the linewidths do not. As discussed in the main text, we speculate that this broadening may be caused by interactions with the electron-hole plasma. Figure~\ref{BayesianFig}(d) shows the energies of the exciton resonances, which follow a ~$n^{-2}$ dependence as expected for Rydberg levels. It also shows the $n$-dependent quantum defects, which start to saturate around $n = 6$, a behavior consistent with previous theoretical studies~\cite{Schone2017}.

\subsection{Calculation of Spectral Fitting Parameters}

From Figure~\ref{BayesianFig}(b), the 10-peak model having the minimum $F$ is the most probable model for Bayesian reconstruction. Therefore, the 10-peak model was used to calculate the fitting parameters. There are some peak-independent parameters and some peak-dependent parameters. The bandgap energy of the system ($E_{g}$), Rydberg binding energy ($Ry$), in-homogeneous broadening ($\sigma$), Urbach absorption coefficient ($a_{0}$), Urbach energy ($E_{u}$) are the peak independent parameters, whereas the quantum defects ($\delta_{n}$), spectral linewidths ($\Gamma_{n}$), Fano asymmetry factors ($q_{n}$), and peak heights ($f_{n}$) are the peak dependent parameters. The fitted parameters are summarized in Table~\ref{peakindependentParams}. and Table~\ref{peakdependentParams}

\begin{table}
    \caption{Peak Independent Parameters}
    \label{peakindependentParams}
    \centering
        \begin{tabular}{|c c c c c|} 
        \hline
       $E_{g}$(eV)  &  $R_{y}$(meV) & $\sigma$ (nm)& $\alpha_{0}$ (unitless)& $E_{u}$ (eV) \\ [0.9ex] 
        \hline\hline
         2.173   & 94.9    &0.02& 0.2864&0.008\\
        \hline
        \end{tabular}
\end{table}

\begin{table}
    \caption{Peak Dependent Parameters}
    \label{peakdependentParams}
    \centering
        \begin{tabular}{|c c c c c|} 
        \hline
       n&$\delta_{n}$  & $\Gamma_{n}$ (nm) &$q_{n}$&$f_{n}$ (a.u.)\\ [0.9ex] 
        \hline\hline
          2 & 0.0096   & 0.66    &3.23& 0.185\\
 3 & 0.043   & 0.249    &4.51& 0.0956\\
 4 & 0.0436   & 0.129    &3.64& 0.0458\\
 5 & 0.0789   & 0.0992    &5.469& 0.0246\\
 6 & 0.0942   & 0.0787    &9.1153& 0.0127\\
 7 & 0.0919   & 0.0611    &9.1415& 0.006\\
 8 & 0.0986   & 0.0414    &9.359& 0.003\\
 9 & 0.0907   & 0.0217    &7.4665& 0.0014\\
 10 & 0.0909   & 0.0103    &7.2898& 0.00076\\
 11 & 0.09   & 0.0051    &6.666& 0.00014\\
        \hline
        \end{tabular}
\end{table}

\newpage

\bibliography{apssamp}

\providecommand{\noopsort}[1]{}\providecommand{\singleletter}[1]{#1}%
\begin{thebibliography}{54}%
\makeatletter
\providecommand \@ifxundefined [1]{%
 \@ifx{#1\undefined}
}%
\providecommand \@ifnum [1]{%
 \ifnum #1\expandafter \@firstoftwo
 \else \expandafter \@secondoftwo
 \fi
}%
\providecommand \@ifx [1]{%
 \ifx #1\expandafter \@firstoftwo
 \else \expandafter \@secondoftwo
 \fi
}%
\providecommand \natexlab [1]{#1}%
\providecommand \enquote  [1]{``#1''}%
\providecommand \bibnamefont  [1]{#1}%
\providecommand \bibfnamefont [1]{#1}%
\providecommand \citenamefont [1]{#1}%
\providecommand \href@noop [0]{\@secondoftwo}%
\providecommand \href [0]{\begingroup \@sanitize@url \@href}%
\providecommand \@href[1]{\@@startlink{#1}\@@href}%
\providecommand \@@href[1]{\endgroup#1\@@endlink}%
\providecommand \@sanitize@url [0]{\catcode `\\12\catcode `\$12\catcode
  `\&12\catcode `\#12\catcode `\^12\catcode `\_12\catcode `\%12\relax}%
\providecommand \@@startlink[1]{}%
\providecommand \@@endlink[0]{}%
\providecommand \url  [0]{\begingroup\@sanitize@url \@url }%
\providecommand \@url [1]{\endgroup\@href {#1}{\urlprefix }}%
\providecommand \urlprefix  [0]{URL }%
\providecommand \Eprint [0]{\href }%
\providecommand \doibase [0]{https://doi.org/}%
\providecommand \selectlanguage [0]{\@gobble}%
\providecommand \bibinfo  [0]{\@secondoftwo}%
\providecommand \bibfield  [0]{\@secondoftwo}%
\providecommand \translation [1]{[#1]}%
\providecommand \BibitemOpen [0]{}%
\providecommand \bibitemStop [0]{}%
\providecommand \bibitemNoStop [0]{.\EOS\space}%
\providecommand \EOS [0]{\spacefactor3000\relax}%
\providecommand \BibitemShut  [1]{\csname bibitem#1\endcsname}%
\let\auto@bib@innerbib\@empty
\bibitem [{\citenamefont {Wang}\ \emph {et~al.}(2020)\citenamefont {Wang},
  \citenamefont {Sciarrino}, \citenamefont {Laing},\ and\ \citenamefont
  {Thompson}}]{Wang2020}%
  \BibitemOpen
  \bibfield  {author} {\bibinfo {author} {\bibfnamefont {J.}~\bibnamefont
  {Wang}}, \bibinfo {author} {\bibfnamefont {F.}~\bibnamefont {Sciarrino}},
  \bibinfo {author} {\bibfnamefont {A.}~\bibnamefont {Laing}},\ and\ \bibinfo
  {author} {\bibfnamefont {M.~G.}\ \bibnamefont {Thompson}},\ }\bibfield
  {title} {\bibinfo {title} {Integrated photonic quantum technologies},\ }\href
  {https://doi.org/10.1038/s41566-019-0532-1} {\bibfield  {journal} {\bibinfo
  {journal} {Nature Photonics}\ }\textbf {\bibinfo {volume} {14}},\ \bibinfo
  {pages} {273} (\bibinfo {year} {2020})}\BibitemShut {NoStop}%
\bibitem [{\citenamefont {Vigliar}\ \emph {et~al.}(2021)\citenamefont
  {Vigliar}, \citenamefont {Paesani}, \citenamefont {Ding}, \citenamefont
  {Adcock}, \citenamefont {Wang}, \citenamefont {Morley-Short}, \citenamefont
  {Bacco}, \citenamefont {Oxenl{\o}we}, \citenamefont {Thompson}, \citenamefont
  {Rarity},\ and\ \citenamefont {Laing}}]{Vigliar2021}%
  \BibitemOpen
  \bibfield  {author} {\bibinfo {author} {\bibfnamefont {C.}~\bibnamefont
  {Vigliar}}, \bibinfo {author} {\bibfnamefont {S.}~\bibnamefont {Paesani}},
  \bibinfo {author} {\bibfnamefont {Y.}~\bibnamefont {Ding}}, \bibinfo {author}
  {\bibfnamefont {J.~C.}\ \bibnamefont {Adcock}}, \bibinfo {author}
  {\bibfnamefont {J.}~\bibnamefont {Wang}}, \bibinfo {author} {\bibfnamefont
  {S.}~\bibnamefont {Morley-Short}}, \bibinfo {author} {\bibfnamefont
  {D.}~\bibnamefont {Bacco}}, \bibinfo {author} {\bibfnamefont {L.~K.}\
  \bibnamefont {Oxenl{\o}we}}, \bibinfo {author} {\bibfnamefont {M.~G.}\
  \bibnamefont {Thompson}}, \bibinfo {author} {\bibfnamefont {J.~G.}\
  \bibnamefont {Rarity}},\ and\ \bibinfo {author} {\bibfnamefont
  {A.}~\bibnamefont {Laing}},\ }\bibfield  {title} {\bibinfo {title}
  {Error-protected qubits in a silicon photonic chip},\ }\href
  {https://doi.org/10.1038/s41567-021-01333-w} {\bibfield  {journal} {\bibinfo
  {journal} {Nature Physics}\ }\textbf {\bibinfo {volume} {17}},\ \bibinfo
  {pages} {1137} (\bibinfo {year} {2021})}\BibitemShut {NoStop}%
\bibitem [{\citenamefont {Bourassa}\ \emph {et~al.}(2021)\citenamefont
  {Bourassa}, \citenamefont {Alexander}, \citenamefont {Vasmer}, \citenamefont
  {Patil}, \citenamefont {Tzitrin}, \citenamefont {Matsuura}, \citenamefont
  {Su}, \citenamefont {Baragiola}, \citenamefont {Guha}, \citenamefont
  {Dauphinais}, \citenamefont {Sabapathy}, \citenamefont {Menicucci},\ and\
  \citenamefont {Dhand}}]{Bourassa2021}%
  \BibitemOpen
  \bibfield  {author} {\bibinfo {author} {\bibfnamefont {J.~E.}\ \bibnamefont
  {Bourassa}}, \bibinfo {author} {\bibfnamefont {R.~N.}\ \bibnamefont
  {Alexander}}, \bibinfo {author} {\bibfnamefont {M.}~\bibnamefont {Vasmer}},
  \bibinfo {author} {\bibfnamefont {A.}~\bibnamefont {Patil}}, \bibinfo
  {author} {\bibfnamefont {I.}~\bibnamefont {Tzitrin}}, \bibinfo {author}
  {\bibfnamefont {T.}~\bibnamefont {Matsuura}}, \bibinfo {author}
  {\bibfnamefont {D.}~\bibnamefont {Su}}, \bibinfo {author} {\bibfnamefont
  {B.~Q.}\ \bibnamefont {Baragiola}}, \bibinfo {author} {\bibfnamefont
  {S.}~\bibnamefont {Guha}}, \bibinfo {author} {\bibfnamefont {G.}~\bibnamefont
  {Dauphinais}}, \bibinfo {author} {\bibfnamefont {K.~K.}\ \bibnamefont
  {Sabapathy}}, \bibinfo {author} {\bibfnamefont {N.~C.}\ \bibnamefont
  {Menicucci}},\ and\ \bibinfo {author} {\bibfnamefont {I.}~\bibnamefont
  {Dhand}},\ }\bibfield  {title} {\bibinfo {title} {Blueprint for a {S}calable
  {P}hotonic {F}ault-{T}olerant {Q}uantum {C}omputer},\ }\href
  {https://doi.org/10.22331/q-2021-02-04-392} {\bibfield  {journal} {\bibinfo
  {journal} {{Quantum}}\ }\textbf {\bibinfo {volume} {5}},\ \bibinfo {pages}
  {392} (\bibinfo {year} {2021})}\BibitemShut {NoStop}%
\bibitem [{\citenamefont {Aßmann}\ and\ \citenamefont
  {Bayer}(2020)}]{Bayer2020}%
  \BibitemOpen
  \bibfield  {author} {\bibinfo {author} {\bibfnamefont {M.}~\bibnamefont
  {Aßmann}}\ and\ \bibinfo {author} {\bibfnamefont {M.}~\bibnamefont
  {Bayer}},\ }\bibfield  {title} {\bibinfo {title} {Semiconductor rydberg
  physics},\ }\href {https://doi.org/10.1002/qute.201900134} {\bibfield
  {journal} {\bibinfo  {journal} {Adv. Quantum Technol.}\ }\textbf {\bibinfo
  {volume} {3}},\ \bibinfo {pages} {1900134} (\bibinfo {year}
  {2020})}\BibitemShut {NoStop}%
\bibitem [{\citenamefont {Rydberg}(1890)}]{Rydberg1890}%
  \BibitemOpen
  \bibfield  {author} {\bibinfo {author} {\bibfnamefont {J.}~\bibnamefont
  {Rydberg}},\ }\bibfield  {title} {\bibinfo {title} {On the structure of the
  line-spectra of the chemical elements},\ }\href
  {https://doi.org/10.1080/14786449008619945} {\bibfield  {journal} {\bibinfo
  {journal} {London, Edinburgh, Dublin Philos. Mag. J. Sci.}\ }\textbf
  {\bibinfo {volume} {29}},\ \bibinfo {pages} {331} (\bibinfo {year} {1890})},\
  \Eprint {https://arxiv.org/abs/https://doi.org/10.1080/14786449008619945}
  {https://doi.org/10.1080/14786449008619945} \BibitemShut {NoStop}%
\bibitem [{\citenamefont {Orfanakis}\ \emph {et~al.}(2021)\citenamefont
  {Orfanakis}, \citenamefont {Rajendran}, \citenamefont {Ohadi}, \citenamefont
  {Zieli\ifmmode \acute{n}\else \'{n}\fi{}ska-Raczy\ifmmode~\acute{n}\else
  \'{n}\fi{}ska}, \citenamefont {Czajkowski}, \citenamefont
  {Karpi\ifmmode~\acute{n}\else \'{n}\fi{}ski},\ and\ \citenamefont
  {Ziemkiewicz}}]{Orfanakis2021}%
  \BibitemOpen
  \bibfield  {author} {\bibinfo {author} {\bibfnamefont {K.}~\bibnamefont
  {Orfanakis}}, \bibinfo {author} {\bibfnamefont {S.~K.}\ \bibnamefont
  {Rajendran}}, \bibinfo {author} {\bibfnamefont {H.}~\bibnamefont {Ohadi}},
  \bibinfo {author} {\bibfnamefont {S.}~\bibnamefont {Zieli\ifmmode
  \acute{n}\else \'{n}\fi{}ska-Raczy\ifmmode~\acute{n}\else \'{n}\fi{}ska}},
  \bibinfo {author} {\bibfnamefont {G.}~\bibnamefont {Czajkowski}}, \bibinfo
  {author} {\bibfnamefont {K.}~\bibnamefont {Karpi\ifmmode~\acute{n}\else
  \'{n}\fi{}ski}},\ and\ \bibinfo {author} {\bibfnamefont {D.}~\bibnamefont
  {Ziemkiewicz}},\ }\bibfield  {title} {\bibinfo {title} {Quantum confined
  rydberg excitons in {Cu}${}_2${O} nanoparticles},\ }\href
  {https://doi.org/10.1103/PhysRevB.103.245426} {\bibfield  {journal} {\bibinfo
   {journal} {Phys. Rev. B}\ }\textbf {\bibinfo {volume} {103}},\ \bibinfo
  {pages} {245426} (\bibinfo {year} {2021})}\BibitemShut {NoStop}%
\bibitem [{\citenamefont {Gallagher}(1994)}]{Gallagher1994}%
  \BibitemOpen
  \bibfield  {author} {\bibinfo {author} {\bibfnamefont {T.~F.}\ \bibnamefont
  {Gallagher}},\ }\href {https://doi.org/10.1017/CBO9780511524530} {\emph
  {\bibinfo {title} {Rydberg Atoms}}},\ Cambridge Monographs on Atomic,
  Molecular and Chemical Physics\ (\bibinfo  {publisher} {Cambridge University
  Press},\ \bibinfo {year} {1994})\BibitemShut {NoStop}%
\bibitem [{\citenamefont {Šibalić}\ and\ \citenamefont
  {Adams}(2018)}]{Sibalic2018}%
  \BibitemOpen
  \bibfield  {author} {\bibinfo {author} {\bibfnamefont {N.}~\bibnamefont
  {Šibalić}}\ and\ \bibinfo {author} {\bibfnamefont {C.~S.}\ \bibnamefont
  {Adams}},\ }\href {https://doi.org/10.1088/978-0-7503-1635-4} {\emph
  {\bibinfo {title} {Rydberg Physics}}},\ 2399-2891\ (\bibinfo  {publisher}
  {IOP Publishing},\ \bibinfo {year} {2018})\BibitemShut {NoStop}%
\bibitem [{\citenamefont {Urban}\ \emph {et~al.}(2009)\citenamefont {Urban},
  \citenamefont {Johnson}, \citenamefont {Henage}, \citenamefont {Isenhower},
  \citenamefont {Yavuz}, \citenamefont {Walker},\ and\ \citenamefont
  {Saffman}}]{Urban2009}%
  \BibitemOpen
  \bibfield  {author} {\bibinfo {author} {\bibfnamefont {E.}~\bibnamefont
  {Urban}}, \bibinfo {author} {\bibfnamefont {T.~A.}\ \bibnamefont {Johnson}},
  \bibinfo {author} {\bibfnamefont {T.}~\bibnamefont {Henage}}, \bibinfo
  {author} {\bibfnamefont {L.}~\bibnamefont {Isenhower}}, \bibinfo {author}
  {\bibfnamefont {D.~D.}\ \bibnamefont {Yavuz}}, \bibinfo {author}
  {\bibfnamefont {T.~G.}\ \bibnamefont {Walker}},\ and\ \bibinfo {author}
  {\bibfnamefont {M.}~\bibnamefont {Saffman}},\ }\bibfield  {title} {\bibinfo
  {title} {Observation of rydberg blockade between two atoms},\ }\href
  {https://doi.org/10.1038/nphys1178} {\bibfield  {journal} {\bibinfo
  {journal} {Nature Physics}\ }\textbf {\bibinfo {volume} {5}},\ \bibinfo
  {pages} {110} (\bibinfo {year} {2009})}\BibitemShut {NoStop}%
\bibitem [{\citenamefont {Gorniaczyk}\ \emph {et~al.}(2014)\citenamefont
  {Gorniaczyk}, \citenamefont {Tresp}, \citenamefont {Schmidt}, \citenamefont
  {Fedder},\ and\ \citenamefont {Hofferberth}}]{Gorniaczyk2014}%
  \BibitemOpen
  \bibfield  {author} {\bibinfo {author} {\bibfnamefont {H.}~\bibnamefont
  {Gorniaczyk}}, \bibinfo {author} {\bibfnamefont {C.}~\bibnamefont {Tresp}},
  \bibinfo {author} {\bibfnamefont {J.}~\bibnamefont {Schmidt}}, \bibinfo
  {author} {\bibfnamefont {H.}~\bibnamefont {Fedder}},\ and\ \bibinfo {author}
  {\bibfnamefont {S.}~\bibnamefont {Hofferberth}},\ }\bibfield  {title}
  {\bibinfo {title} {Single-photon transistor mediated by interstate rydberg
  interactions},\ }\href {https://doi.org/10.1103/PhysRevLett.113.053601}
  {\bibfield  {journal} {\bibinfo  {journal} {Phys. Rev. Lett.}\ }\textbf
  {\bibinfo {volume} {113}},\ \bibinfo {pages} {053601} (\bibinfo {year}
  {2014})}\BibitemShut {NoStop}%
\bibitem [{\citenamefont {Tiarks}\ \emph {et~al.}(2014)\citenamefont {Tiarks},
  \citenamefont {Baur}, \citenamefont {Schneider}, \citenamefont {D\"urr},\
  and\ \citenamefont {Rempe}}]{Tiarks2014}%
  \BibitemOpen
  \bibfield  {author} {\bibinfo {author} {\bibfnamefont {D.}~\bibnamefont
  {Tiarks}}, \bibinfo {author} {\bibfnamefont {S.}~\bibnamefont {Baur}},
  \bibinfo {author} {\bibfnamefont {K.}~\bibnamefont {Schneider}}, \bibinfo
  {author} {\bibfnamefont {S.}~\bibnamefont {D\"urr}},\ and\ \bibinfo {author}
  {\bibfnamefont {G.}~\bibnamefont {Rempe}},\ }\bibfield  {title} {\bibinfo
  {title} {Single-photon transistor using a f\"orster resonance},\ }\href
  {https://doi.org/10.1103/PhysRevLett.113.053602} {\bibfield  {journal}
  {\bibinfo  {journal} {Phys. Rev. Lett.}\ }\textbf {\bibinfo {volume} {113}},\
  \bibinfo {pages} {053602} (\bibinfo {year} {2014})}\BibitemShut {NoStop}%
\bibitem [{\citenamefont {Levine}\ \emph {et~al.}(2018)\citenamefont {Levine},
  \citenamefont {Keesling}, \citenamefont {Omran}, \citenamefont {Bernien},
  \citenamefont {Schwartz}, \citenamefont {Zibrov}, \citenamefont {Endres},
  \citenamefont {Greiner}, \citenamefont {Vuleti\ifmmode~\acute{c}\else
  \'{c}\fi{}},\ and\ \citenamefont {Lukin}}]{Levine2018}%
  \BibitemOpen
  \bibfield  {author} {\bibinfo {author} {\bibfnamefont {H.}~\bibnamefont
  {Levine}}, \bibinfo {author} {\bibfnamefont {A.}~\bibnamefont {Keesling}},
  \bibinfo {author} {\bibfnamefont {A.}~\bibnamefont {Omran}}, \bibinfo
  {author} {\bibfnamefont {H.}~\bibnamefont {Bernien}}, \bibinfo {author}
  {\bibfnamefont {S.}~\bibnamefont {Schwartz}}, \bibinfo {author}
  {\bibfnamefont {A.~S.}\ \bibnamefont {Zibrov}}, \bibinfo {author}
  {\bibfnamefont {M.}~\bibnamefont {Endres}}, \bibinfo {author} {\bibfnamefont
  {M.}~\bibnamefont {Greiner}}, \bibinfo {author} {\bibfnamefont
  {V.}~\bibnamefont {Vuleti\ifmmode~\acute{c}\else \'{c}\fi{}}},\ and\ \bibinfo
  {author} {\bibfnamefont {M.~D.}\ \bibnamefont {Lukin}},\ }\bibfield  {title}
  {\bibinfo {title} {High-fidelity control and entanglement of rydberg-atom
  qubits},\ }\href {https://doi.org/10.1103/PhysRevLett.121.123603} {\bibfield
  {journal} {\bibinfo  {journal} {Phys. Rev. Lett.}\ }\textbf {\bibinfo
  {volume} {121}},\ \bibinfo {pages} {123603} (\bibinfo {year}
  {2018})}\BibitemShut {NoStop}%
\bibitem [{\citenamefont {Saffman}(2016)}]{Saffman2016}%
  \BibitemOpen
  \bibfield  {author} {\bibinfo {author} {\bibfnamefont {M.}~\bibnamefont
  {Saffman}},\ }\bibfield  {title} {\bibinfo {title} {Quantum computing with
  atomic qubits and rydberg interactions: progress and challenges},\ }\href
  {https://doi.org/10.1088/0953-4075/49/20/202001} {\bibfield  {journal}
  {\bibinfo  {journal} {Journal of Physics B: Atomic, Molecular and Optical
  Physics}\ }\textbf {\bibinfo {volume} {49}},\ \bibinfo {pages} {202001}
  (\bibinfo {year} {2016})}\BibitemShut {NoStop}%
\bibitem [{\citenamefont {Scholl}\ \emph {et~al.}(2021)\citenamefont {Scholl},
  \citenamefont {Schuler}, \citenamefont {Williams}, \citenamefont
  {Eberharter}, \citenamefont {Barredo}, \citenamefont {Schymik}, \citenamefont
  {Lienhard}, \citenamefont {Henry}, \citenamefont {Lang}, \citenamefont
  {Lahaye}, \citenamefont {L{\"a}uchli},\ and\ \citenamefont
  {Browaeys}}]{Scholl2021}%
  \BibitemOpen
  \bibfield  {author} {\bibinfo {author} {\bibfnamefont {P.}~\bibnamefont
  {Scholl}}, \bibinfo {author} {\bibfnamefont {M.}~\bibnamefont {Schuler}},
  \bibinfo {author} {\bibfnamefont {H.~J.}\ \bibnamefont {Williams}}, \bibinfo
  {author} {\bibfnamefont {A.~A.}\ \bibnamefont {Eberharter}}, \bibinfo
  {author} {\bibfnamefont {D.}~\bibnamefont {Barredo}}, \bibinfo {author}
  {\bibfnamefont {K.-N.}\ \bibnamefont {Schymik}}, \bibinfo {author}
  {\bibfnamefont {V.}~\bibnamefont {Lienhard}}, \bibinfo {author}
  {\bibfnamefont {L.-P.}\ \bibnamefont {Henry}}, \bibinfo {author}
  {\bibfnamefont {T.~C.}\ \bibnamefont {Lang}}, \bibinfo {author}
  {\bibfnamefont {T.}~\bibnamefont {Lahaye}}, \bibinfo {author} {\bibfnamefont
  {A.~M.}\ \bibnamefont {L{\"a}uchli}},\ and\ \bibinfo {author} {\bibfnamefont
  {A.}~\bibnamefont {Browaeys}},\ }\bibfield  {title} {\bibinfo {title}
  {Quantum simulation of 2d antiferromagnets with hundreds of rydberg atoms},\
  }\href {https://doi.org/10.1038/s41586-021-03585-1} {\bibfield  {journal}
  {\bibinfo  {journal} {Nature}\ }\textbf {\bibinfo {volume} {595}},\ \bibinfo
  {pages} {233} (\bibinfo {year} {2021})}\BibitemShut {NoStop}%
\bibitem [{\citenamefont {Ebadi}\ \emph {et~al.}(2021)\citenamefont {Ebadi},
  \citenamefont {Wang}, \citenamefont {Levine}, \citenamefont {Keesling},
  \citenamefont {Semeghini}, \citenamefont {Omran}, \citenamefont {Bluvstein},
  \citenamefont {Samajdar}, \citenamefont {Pichler}, \citenamefont {Ho},
  \citenamefont {Choi}, \citenamefont {Sachdev}, \citenamefont {Greiner},
  \citenamefont {Vuleti{\'{c}}},\ and\ \citenamefont {Lukin}}]{Ebadi2021}%
  \BibitemOpen
  \bibfield  {author} {\bibinfo {author} {\bibfnamefont {S.}~\bibnamefont
  {Ebadi}}, \bibinfo {author} {\bibfnamefont {T.~T.}\ \bibnamefont {Wang}},
  \bibinfo {author} {\bibfnamefont {H.}~\bibnamefont {Levine}}, \bibinfo
  {author} {\bibfnamefont {A.}~\bibnamefont {Keesling}}, \bibinfo {author}
  {\bibfnamefont {G.}~\bibnamefont {Semeghini}}, \bibinfo {author}
  {\bibfnamefont {A.}~\bibnamefont {Omran}}, \bibinfo {author} {\bibfnamefont
  {D.}~\bibnamefont {Bluvstein}}, \bibinfo {author} {\bibfnamefont
  {R.}~\bibnamefont {Samajdar}}, \bibinfo {author} {\bibfnamefont
  {H.}~\bibnamefont {Pichler}}, \bibinfo {author} {\bibfnamefont {W.~W.}\
  \bibnamefont {Ho}}, \bibinfo {author} {\bibfnamefont {S.}~\bibnamefont
  {Choi}}, \bibinfo {author} {\bibfnamefont {S.}~\bibnamefont {Sachdev}},
  \bibinfo {author} {\bibfnamefont {M.}~\bibnamefont {Greiner}}, \bibinfo
  {author} {\bibfnamefont {V.}~\bibnamefont {Vuleti{\'{c}}}},\ and\ \bibinfo
  {author} {\bibfnamefont {M.~D.}\ \bibnamefont {Lukin}},\ }\bibfield  {title}
  {\bibinfo {title} {Quantum phases of matter on a 256-atom programmable
  quantum simulator},\ }\href {https://doi.org/10.1038/s41586-021-03582-4}
  {\bibfield  {journal} {\bibinfo  {journal} {Nature}\ }\textbf {\bibinfo
  {volume} {595}},\ \bibinfo {pages} {227} (\bibinfo {year}
  {2021})}\BibitemShut {NoStop}%
\bibitem [{\citenamefont {Walther}\ \emph {et~al.}(2018)\citenamefont
  {Walther}, \citenamefont {Johne},\ and\ \citenamefont {Pohl}}]{Walther2018}%
  \BibitemOpen
  \bibfield  {author} {\bibinfo {author} {\bibfnamefont {V.}~\bibnamefont
  {Walther}}, \bibinfo {author} {\bibfnamefont {R.}~\bibnamefont {Johne}},\
  and\ \bibinfo {author} {\bibfnamefont {T.}~\bibnamefont {Pohl}},\ }\bibfield
  {title} {\bibinfo {title} {Giant optical nonlinearities from rydberg excitons
  in semiconductor microcavities},\ }\href
  {https://doi.org/10.1038/s41467-018-03742-7} {\bibfield  {journal} {\bibinfo
  {journal} {Nature Communications}\ }\textbf {\bibinfo {volume} {9}},\
  \bibinfo {pages} {1309} (\bibinfo {year} {2018})}\BibitemShut {NoStop}%
\bibitem [{\citenamefont {Zhao}\ \emph {et~al.}(2021)\citenamefont {Zhao},
  \citenamefont {Shang}, \citenamefont {Li}, \citenamefont {Liang},
  \citenamefont {Li},\ and\ \citenamefont {Zhang}}]{Zhao2021}%
  \BibitemOpen
  \bibfield  {author} {\bibinfo {author} {\bibfnamefont {L.}~\bibnamefont
  {Zhao}}, \bibinfo {author} {\bibfnamefont {Q.}~\bibnamefont {Shang}},
  \bibinfo {author} {\bibfnamefont {M.}~\bibnamefont {Li}}, \bibinfo {author}
  {\bibfnamefont {Y.}~\bibnamefont {Liang}}, \bibinfo {author} {\bibfnamefont
  {C.}~\bibnamefont {Li}},\ and\ \bibinfo {author} {\bibfnamefont
  {Q.}~\bibnamefont {Zhang}},\ }\bibfield  {title} {\bibinfo {title} {Strong
  exciton-photon interaction and lasing of two-dimensional transition metal
  dichalcogenide semiconductors},\ }\href
  {https://doi.org/10.1007/s12274-020-3073-5} {\bibfield  {journal} {\bibinfo
  {journal} {Nano Research}\ }\textbf {\bibinfo {volume} {14}},\ \bibinfo
  {pages} {1937} (\bibinfo {year} {2021})}\BibitemShut {NoStop}%
\bibitem [{\citenamefont {Su}\ \emph {et~al.}(2021)\citenamefont {Su},
  \citenamefont {Fieramosca}, \citenamefont {Zhang}, \citenamefont {Nguyen},
  \citenamefont {Deleporte}, \citenamefont {Chen}, \citenamefont {Sanvitto},
  \citenamefont {Liew},\ and\ \citenamefont {Xiong}}]{Su2021}%
  \BibitemOpen
  \bibfield  {author} {\bibinfo {author} {\bibfnamefont {R.}~\bibnamefont
  {Su}}, \bibinfo {author} {\bibfnamefont {A.}~\bibnamefont {Fieramosca}},
  \bibinfo {author} {\bibfnamefont {Q.}~\bibnamefont {Zhang}}, \bibinfo
  {author} {\bibfnamefont {H.~S.}\ \bibnamefont {Nguyen}}, \bibinfo {author}
  {\bibfnamefont {E.}~\bibnamefont {Deleporte}}, \bibinfo {author}
  {\bibfnamefont {Z.}~\bibnamefont {Chen}}, \bibinfo {author} {\bibfnamefont
  {D.}~\bibnamefont {Sanvitto}}, \bibinfo {author} {\bibfnamefont {T.~C.~H.}\
  \bibnamefont {Liew}},\ and\ \bibinfo {author} {\bibfnamefont
  {Q.}~\bibnamefont {Xiong}},\ }\bibfield  {title} {\bibinfo {title}
  {Perovskite semiconductors for room-temperature exciton-polaritonics},\
  }\href {https://doi.org/10.1038/s41563-021-01035-x} {\bibfield  {journal}
  {\bibinfo  {journal} {Nature Materials}\ }\textbf {\bibinfo {volume} {20}},\
  \bibinfo {pages} {1315} (\bibinfo {year} {2021})}\BibitemShut {NoStop}%
\bibitem [{\citenamefont {Elliott}(1961)}]{Elliott1961}%
  \BibitemOpen
  \bibfield  {author} {\bibinfo {author} {\bibfnamefont {R.~J.}\ \bibnamefont
  {Elliott}},\ }\bibfield  {title} {\bibinfo {title} {Symmetry of excitons in
  {Cu}${}_2${O}},\ }\href {https://doi.org/10.1103/PhysRev.124.340} {\bibfield
  {journal} {\bibinfo  {journal} {Phys. Rev.}\ }\textbf {\bibinfo {volume}
  {124}},\ \bibinfo {pages} {340} (\bibinfo {year} {1961})}\BibitemShut
  {NoStop}%
\bibitem [{\citenamefont {Otter}(2007)}]{Otter2007}%
  \BibitemOpen
  \bibfield  {author} {\bibinfo {author} {\bibfnamefont {M.}~\bibnamefont
  {Otter}},\ }\href
  {https://www.rug.nl/research/zernike/education/topmasternanoscience/ns201otter.pdf}
  {\bibinfo {title} {Lifetime of paraexcitons in cuprous oxide}} (\bibinfo
  {year} {2007})\BibitemShut {NoStop}%
\bibitem [{\citenamefont {Zhu}\ \emph {et~al.}(2015)\citenamefont {Zhu},
  \citenamefont {Chen},\ and\ \citenamefont {Cui}}]{Zhu2015}%
  \BibitemOpen
  \bibfield  {author} {\bibinfo {author} {\bibfnamefont {B.}~\bibnamefont
  {Zhu}}, \bibinfo {author} {\bibfnamefont {X.}~\bibnamefont {Chen}},\ and\
  \bibinfo {author} {\bibfnamefont {X.}~\bibnamefont {Cui}},\ }\bibfield
  {title} {\bibinfo {title} {Exciton binding energy of monolayer {WS}${}_2$},\
  }\href {https://doi.org/10.1038/srep09218} {\bibfield  {journal} {\bibinfo
  {journal} {Scientific Reports}\ }\textbf {\bibinfo {volume} {5}},\ \bibinfo
  {pages} {9218} (\bibinfo {year} {2015})}\BibitemShut {NoStop}%
\bibitem [{\citenamefont {Park}\ \emph {et~al.}(2018)\citenamefont {Park},
  \citenamefont {Mutz}, \citenamefont {Schultz}, \citenamefont {Blumstengel},
  \citenamefont {Han}, \citenamefont {Aljarb}, \citenamefont {Li},
  \citenamefont {List-Kratochvil}, \citenamefont {Amsalem},\ and\ \citenamefont
  {Koch}}]{Park2018}%
  \BibitemOpen
  \bibfield  {author} {\bibinfo {author} {\bibfnamefont {S.}~\bibnamefont
  {Park}}, \bibinfo {author} {\bibfnamefont {N.}~\bibnamefont {Mutz}}, \bibinfo
  {author} {\bibfnamefont {T.}~\bibnamefont {Schultz}}, \bibinfo {author}
  {\bibfnamefont {S.}~\bibnamefont {Blumstengel}}, \bibinfo {author}
  {\bibfnamefont {A.}~\bibnamefont {Han}}, \bibinfo {author} {\bibfnamefont
  {A.}~\bibnamefont {Aljarb}}, \bibinfo {author} {\bibfnamefont {L.-J.}\
  \bibnamefont {Li}}, \bibinfo {author} {\bibfnamefont {E.~J.~W.}\ \bibnamefont
  {List-Kratochvil}}, \bibinfo {author} {\bibfnamefont {P.}~\bibnamefont
  {Amsalem}},\ and\ \bibinfo {author} {\bibfnamefont {N.}~\bibnamefont
  {Koch}},\ }\bibfield  {title} {\bibinfo {title} {Direct determination of
  monolayer {MoS}${}_2$ and {WSe}${}_2$ exciton binding energies on insulating
  and metallic substrates},\ }\href {https://doi.org/10.1088/2053-1583/aaa4ca}
  {\bibfield  {journal} {\bibinfo  {journal} {2D Materials}\ }\textbf {\bibinfo
  {volume} {5}},\ \bibinfo {pages} {025003} (\bibinfo {year}
  {2018})}\BibitemShut {NoStop}%
\bibitem [{\citenamefont {Robert}\ \emph {et~al.}(2016)\citenamefont {Robert},
  \citenamefont {Lagarde}, \citenamefont {Cadiz}, \citenamefont {Wang},
  \citenamefont {Lassagne}, \citenamefont {Amand}, \citenamefont {Balocchi},
  \citenamefont {Renucci}, \citenamefont {Tongay}, \citenamefont {Urbaszek},\
  and\ \citenamefont {Marie}}]{Robert2016}%
  \BibitemOpen
  \bibfield  {author} {\bibinfo {author} {\bibfnamefont {C.}~\bibnamefont
  {Robert}}, \bibinfo {author} {\bibfnamefont {D.}~\bibnamefont {Lagarde}},
  \bibinfo {author} {\bibfnamefont {F.}~\bibnamefont {Cadiz}}, \bibinfo
  {author} {\bibfnamefont {G.}~\bibnamefont {Wang}}, \bibinfo {author}
  {\bibfnamefont {B.}~\bibnamefont {Lassagne}}, \bibinfo {author}
  {\bibfnamefont {T.}~\bibnamefont {Amand}}, \bibinfo {author} {\bibfnamefont
  {A.}~\bibnamefont {Balocchi}}, \bibinfo {author} {\bibfnamefont
  {P.}~\bibnamefont {Renucci}}, \bibinfo {author} {\bibfnamefont
  {S.}~\bibnamefont {Tongay}}, \bibinfo {author} {\bibfnamefont
  {B.}~\bibnamefont {Urbaszek}},\ and\ \bibinfo {author} {\bibfnamefont
  {X.}~\bibnamefont {Marie}},\ }\bibfield  {title} {\bibinfo {title} {Exciton
  radiative lifetime in transition metal dichalcogenide monolayers},\ }\href
  {https://doi.org/10.1103/PhysRevB.93.205423} {\bibfield  {journal} {\bibinfo
  {journal} {Phys. Rev. B}\ }\textbf {\bibinfo {volume} {93}},\ \bibinfo
  {pages} {205423} (\bibinfo {year} {2016})}\BibitemShut {NoStop}%
\bibitem [{\citenamefont {Kazimierczuk}\ \emph {et~al.}(2014)\citenamefont
  {Kazimierczuk}, \citenamefont {Fr{\"o}hlich}, \citenamefont {Scheel},
  \citenamefont {Stolz},\ and\ \citenamefont {Bayer}}]{Kazimierczuk2014}%
  \BibitemOpen
  \bibfield  {author} {\bibinfo {author} {\bibfnamefont {T.}~\bibnamefont
  {Kazimierczuk}}, \bibinfo {author} {\bibfnamefont {D.}~\bibnamefont
  {Fr{\"o}hlich}}, \bibinfo {author} {\bibfnamefont {S.}~\bibnamefont
  {Scheel}}, \bibinfo {author} {\bibfnamefont {H.}~\bibnamefont {Stolz}},\ and\
  \bibinfo {author} {\bibfnamefont {M.}~\bibnamefont {Bayer}},\ }\bibfield
  {title} {\bibinfo {title} {Giant rydberg excitons in the copper oxide
  {Cu}${}_2${O}},\ }\href {https://doi.org/10.1038/nature13832} {\bibfield
  {journal} {\bibinfo  {journal} {Nature}\ }\textbf {\bibinfo {volume} {514}},\
  \bibinfo {pages} {343} (\bibinfo {year} {2014})}\BibitemShut {NoStop}%
\bibitem [{\citenamefont {Takahata}\ and\ \citenamefont
  {Naka}(2018)}]{Takahata2018}%
  \BibitemOpen
  \bibfield  {author} {\bibinfo {author} {\bibfnamefont {M.}~\bibnamefont
  {Takahata}}\ and\ \bibinfo {author} {\bibfnamefont {N.}~\bibnamefont
  {Naka}},\ }\bibfield  {title} {\bibinfo {title} {Photoluminescence properties
  of the entire excitonic series in {Cu}${}_2${O}},\ }\href
  {https://doi.org/10.1103/PhysRevB.98.195205} {\bibfield  {journal} {\bibinfo
  {journal} {Phys. Rev. B}\ }\textbf {\bibinfo {volume} {98}},\ \bibinfo
  {pages} {195205} (\bibinfo {year} {2018})}\BibitemShut {NoStop}%
\bibitem [{\citenamefont {Mund}\ \emph {et~al.}(2018)\citenamefont {Mund},
  \citenamefont {Fr\"ohlich}, \citenamefont {Yakovlev},\ and\ \citenamefont
  {Bayer}}]{Mund2018}%
  \BibitemOpen
  \bibfield  {author} {\bibinfo {author} {\bibfnamefont {J.}~\bibnamefont
  {Mund}}, \bibinfo {author} {\bibfnamefont {D.}~\bibnamefont {Fr\"ohlich}},
  \bibinfo {author} {\bibfnamefont {D.~R.}\ \bibnamefont {Yakovlev}},\ and\
  \bibinfo {author} {\bibfnamefont {M.}~\bibnamefont {Bayer}},\ }\bibfield
  {title} {\bibinfo {title} {High-resolution second harmonic generation
  spectroscopy with femtosecond laser pulses on excitons in {Cu}${}_2${O}},\
  }\href {https://doi.org/10.1103/PhysRevB.98.085203} {\bibfield  {journal}
  {\bibinfo  {journal} {Phys. Rev. B}\ }\textbf {\bibinfo {volume} {98}},\
  \bibinfo {pages} {085203} (\bibinfo {year} {2018})}\BibitemShut {NoStop}%
\bibitem [{\citenamefont {Kang}\ \emph {et~al.}(2021)\citenamefont {Kang},
  \citenamefont {Gross}, \citenamefont {Yang}, \citenamefont {Morita},
  \citenamefont {Choi}, \citenamefont {Yoshioka},\ and\ \citenamefont
  {Kim}}]{Kang2021}%
  \BibitemOpen
  \bibfield  {author} {\bibinfo {author} {\bibfnamefont {D.~D.}\ \bibnamefont
  {Kang}}, \bibinfo {author} {\bibfnamefont {A.}~\bibnamefont {Gross}},
  \bibinfo {author} {\bibfnamefont {H.}~\bibnamefont {Yang}}, \bibinfo {author}
  {\bibfnamefont {Y.}~\bibnamefont {Morita}}, \bibinfo {author} {\bibfnamefont
  {K.~S.}\ \bibnamefont {Choi}}, \bibinfo {author} {\bibfnamefont
  {K.}~\bibnamefont {Yoshioka}},\ and\ \bibinfo {author} {\bibfnamefont
  {N.~Y.}\ \bibnamefont {Kim}},\ }\bibfield  {title} {\bibinfo {title}
  {Temperature study of rydberg exciton optical properties in {Cu}${}_2${O}},\
  }\href {https://doi.org/10.1103/PhysRevB.103.205203} {\bibfield  {journal}
  {\bibinfo  {journal} {Phys. Rev. B}\ }\textbf {\bibinfo {volume} {103}},\
  \bibinfo {pages} {205203} (\bibinfo {year} {2021})}\BibitemShut {NoStop}%
\bibitem [{\citenamefont {Versteegh}\ \emph {et~al.}(2021)\citenamefont
  {Versteegh}, \citenamefont {Steinhauer}, \citenamefont {Bajo}, \citenamefont
  {Lettner}, \citenamefont {Soro}, \citenamefont {Romanova}, \citenamefont
  {Gyger}, \citenamefont {Schweickert}, \citenamefont {Mysyrowicz},\ and\
  \citenamefont {Zwiller}}]{Versteegh2021}%
  \BibitemOpen
  \bibfield  {author} {\bibinfo {author} {\bibfnamefont {M.~A.~M.}\
  \bibnamefont {Versteegh}}, \bibinfo {author} {\bibfnamefont {S.}~\bibnamefont
  {Steinhauer}}, \bibinfo {author} {\bibfnamefont {J.}~\bibnamefont {Bajo}},
  \bibinfo {author} {\bibfnamefont {T.}~\bibnamefont {Lettner}}, \bibinfo
  {author} {\bibfnamefont {A.}~\bibnamefont {Soro}}, \bibinfo {author}
  {\bibfnamefont {A.}~\bibnamefont {Romanova}}, \bibinfo {author}
  {\bibfnamefont {S.}~\bibnamefont {Gyger}}, \bibinfo {author} {\bibfnamefont
  {L.}~\bibnamefont {Schweickert}}, \bibinfo {author} {\bibfnamefont
  {A.}~\bibnamefont {Mysyrowicz}},\ and\ \bibinfo {author} {\bibfnamefont
  {V.}~\bibnamefont {Zwiller}},\ }\bibfield  {title} {\bibinfo {title} {Giant
  rydberg excitons in {Cu}${}_2${O} probed by photoluminescence excitation
  spectroscopy},\ }\href {https://doi.org/10.1103/PhysRevB.104.245206}
  {\bibfield  {journal} {\bibinfo  {journal} {Phys. Rev. B}\ }\textbf {\bibinfo
  {volume} {104}},\ \bibinfo {pages} {245206} (\bibinfo {year}
  {2021})}\BibitemShut {NoStop}%
\bibitem [{\citenamefont {Gallagher}\ \emph {et~al.}(2022)\citenamefont
  {Gallagher}, \citenamefont {Rogers}, \citenamefont {Pritchett}, \citenamefont
  {Mistry}, \citenamefont {Pizzey}, \citenamefont {Adams}, \citenamefont
  {Jones}, \citenamefont {Gr\"unwald}, \citenamefont {Walther}, \citenamefont
  {Hodges}, \citenamefont {Langbein},\ and\ \citenamefont
  {Lynch}}]{Gallagher2022}%
  \BibitemOpen
  \bibfield  {author} {\bibinfo {author} {\bibfnamefont {L.~A.~P.}\
  \bibnamefont {Gallagher}}, \bibinfo {author} {\bibfnamefont {J.~P.}\
  \bibnamefont {Rogers}}, \bibinfo {author} {\bibfnamefont {J.~D.}\
  \bibnamefont {Pritchett}}, \bibinfo {author} {\bibfnamefont {R.~A.}\
  \bibnamefont {Mistry}}, \bibinfo {author} {\bibfnamefont {D.}~\bibnamefont
  {Pizzey}}, \bibinfo {author} {\bibfnamefont {C.~S.}\ \bibnamefont {Adams}},
  \bibinfo {author} {\bibfnamefont {M.~P.~A.}\ \bibnamefont {Jones}}, \bibinfo
  {author} {\bibfnamefont {P.}~\bibnamefont {Gr\"unwald}}, \bibinfo {author}
  {\bibfnamefont {V.}~\bibnamefont {Walther}}, \bibinfo {author} {\bibfnamefont
  {C.}~\bibnamefont {Hodges}}, \bibinfo {author} {\bibfnamefont
  {W.}~\bibnamefont {Langbein}},\ and\ \bibinfo {author} {\bibfnamefont
  {S.~A.}\ \bibnamefont {Lynch}},\ }\bibfield  {title} {\bibinfo {title}
  {Microwave-optical coupling via rydberg excitons in cuprous oxide},\ }\href
  {https://doi.org/10.1103/PhysRevResearch.4.013031} {\bibfield  {journal}
  {\bibinfo  {journal} {Phys. Rev. Research}\ }\textbf {\bibinfo {volume}
  {4}},\ \bibinfo {pages} {013031} (\bibinfo {year} {2022})}\BibitemShut
  {NoStop}%
\bibitem [{\citenamefont {Orfanakis}\ \emph {et~al.}(2022)\citenamefont
  {Orfanakis}, \citenamefont {Rajendran}, \citenamefont {Walther},
  \citenamefont {Volz}, \citenamefont {Pohl},\ and\ \citenamefont
  {Ohadi}}]{Orfanakis2022}%
  \BibitemOpen
  \bibfield  {author} {\bibinfo {author} {\bibfnamefont {K.}~\bibnamefont
  {Orfanakis}}, \bibinfo {author} {\bibfnamefont {S.~K.}\ \bibnamefont
  {Rajendran}}, \bibinfo {author} {\bibfnamefont {V.}~\bibnamefont {Walther}},
  \bibinfo {author} {\bibfnamefont {T.}~\bibnamefont {Volz}}, \bibinfo {author}
  {\bibfnamefont {T.}~\bibnamefont {Pohl}},\ and\ \bibinfo {author}
  {\bibfnamefont {H.}~\bibnamefont {Ohadi}},\ }\bibfield  {title} {\bibinfo
  {title} {Rydberg exciton--polaritons in a {Cu}${}_2${O} microcavity},\ }\href
  {https://doi.org/10.1038/s41563-022-01230-4} {\bibfield  {journal} {\bibinfo
  {journal} {Nature Materials}\ }\textbf {\bibinfo {volume} {21}},\ \bibinfo
  {pages} {767} (\bibinfo {year} {2022})}\BibitemShut {NoStop}%
\bibitem [{\citenamefont {Steinhauer}\ \emph {et~al.}(2020)\citenamefont
  {Steinhauer}, \citenamefont {Versteegh}, \citenamefont {Gyger}, \citenamefont
  {Elshaari}, \citenamefont {Kunert}, \citenamefont {Mysyrowicz},\ and\
  \citenamefont {Zwiller}}]{Steinhauer2020}%
  \BibitemOpen
  \bibfield  {author} {\bibinfo {author} {\bibfnamefont {S.}~\bibnamefont
  {Steinhauer}}, \bibinfo {author} {\bibfnamefont {M.}~\bibnamefont
  {Versteegh}}, \bibinfo {author} {\bibfnamefont {S.}~\bibnamefont {Gyger}},
  \bibinfo {author} {\bibfnamefont {A.}~\bibnamefont {Elshaari}}, \bibinfo
  {author} {\bibfnamefont {B.}~\bibnamefont {Kunert}}, \bibinfo {author}
  {\bibfnamefont {A.}~\bibnamefont {Mysyrowicz}},\ and\ \bibinfo {author}
  {\bibfnamefont {V.}~\bibnamefont {Zwiller}},\ }\bibfield  {title} {\bibinfo
  {title} {Rydberg excitons in {Cu}${}_2${O} microcrystals grown on a silicon
  platform},\ }\bibfield  {journal} {\bibinfo  {journal} {Comm. Mat.}\ }\textbf
  {\bibinfo {volume} {1}},\ \href {https://doi.org/10.1038/s43246-020-0013-6}
  {10.1038/s43246-020-0013-6} (\bibinfo {year} {2020})\BibitemShut {NoStop}%
\bibitem [{\citenamefont {Lynch}\ \emph {et~al.}(2021)\citenamefont {Lynch},
  \citenamefont {Hodges}, \citenamefont {Mandal}, \citenamefont {Langbein},
  \citenamefont {Singh}, \citenamefont {Gallagher}, \citenamefont {Pritchett},
  \citenamefont {Pizzey}, \citenamefont {Rogers}, \citenamefont {Adams},\ and\
  \citenamefont {Jones}}]{Lynch2021}%
  \BibitemOpen
  \bibfield  {author} {\bibinfo {author} {\bibfnamefont {S.~A.}\ \bibnamefont
  {Lynch}}, \bibinfo {author} {\bibfnamefont {C.}~\bibnamefont {Hodges}},
  \bibinfo {author} {\bibfnamefont {S.}~\bibnamefont {Mandal}}, \bibinfo
  {author} {\bibfnamefont {W.}~\bibnamefont {Langbein}}, \bibinfo {author}
  {\bibfnamefont {R.~P.}\ \bibnamefont {Singh}}, \bibinfo {author}
  {\bibfnamefont {L.~A.~P.}\ \bibnamefont {Gallagher}}, \bibinfo {author}
  {\bibfnamefont {J.~D.}\ \bibnamefont {Pritchett}}, \bibinfo {author}
  {\bibfnamefont {D.}~\bibnamefont {Pizzey}}, \bibinfo {author} {\bibfnamefont
  {J.~P.}\ \bibnamefont {Rogers}}, \bibinfo {author} {\bibfnamefont {C.~S.}\
  \bibnamefont {Adams}},\ and\ \bibinfo {author} {\bibfnamefont {M.~P.~A.}\
  \bibnamefont {Jones}},\ }\bibfield  {title} {\bibinfo {title} {Rydberg
  excitons in synthetic cuprous oxide {Cu}${}_2${O}},\ }\href
  {https://doi.org/10.1103/PhysRevMaterials.5.084602} {\bibfield  {journal}
  {\bibinfo  {journal} {Phys. Rev. Materials}\ }\textbf {\bibinfo {volume}
  {5}},\ \bibinfo {pages} {084602} (\bibinfo {year} {2021})}\BibitemShut
  {NoStop}%
\bibitem [{\citenamefont {Khazali}\ \emph {et~al.}(2017)\citenamefont
  {Khazali}, \citenamefont {Heshami},\ and\ \citenamefont
  {Simon}}]{Khazali2017}%
  \BibitemOpen
  \bibfield  {author} {\bibinfo {author} {\bibfnamefont {M.}~\bibnamefont
  {Khazali}}, \bibinfo {author} {\bibfnamefont {K.}~\bibnamefont {Heshami}},\
  and\ \bibinfo {author} {\bibfnamefont {C.}~\bibnamefont {Simon}},\ }\bibfield
   {title} {\bibinfo {title} {Single-photon source based on rydberg exciton
  blockade},\ }\href {https://doi.org/10.1088/1361-6455/aa8d7c} {\bibfield
  {journal} {\bibinfo  {journal} {Journal of Physics B: Atomic, Molecular and
  Optical Physics}\ }\textbf {\bibinfo {volume} {50}},\ \bibinfo {pages}
  {215301} (\bibinfo {year} {2017})}\BibitemShut {NoStop}%
\bibitem [{\citenamefont {Walther}\ \emph {et~al.}(2022)\citenamefont
  {Walther}, \citenamefont {Zhang}, \citenamefont {Yelin},\ and\ \citenamefont
  {Pohl}}]{Walther2022}%
  \BibitemOpen
  \bibfield  {author} {\bibinfo {author} {\bibfnamefont {V.}~\bibnamefont
  {Walther}}, \bibinfo {author} {\bibfnamefont {L.}~\bibnamefont {Zhang}},
  \bibinfo {author} {\bibfnamefont {S.~F.}\ \bibnamefont {Yelin}},\ and\
  \bibinfo {author} {\bibfnamefont {T.}~\bibnamefont {Pohl}},\ }\bibfield
  {title} {\bibinfo {title} {Nonclassical light from finite-range interactions
  in a two-dimensional quantum mirror},\ }\href
  {https://doi.org/10.1103/PhysRevB.105.075307} {\bibfield  {journal} {\bibinfo
   {journal} {Phys. Rev. B}\ }\textbf {\bibinfo {volume} {105}},\ \bibinfo
  {pages} {075307} (\bibinfo {year} {2022})}\BibitemShut {NoStop}%
\bibitem [{\citenamefont {Elliott}(1957)}]{Elliott1957}%
  \BibitemOpen
  \bibfield  {author} {\bibinfo {author} {\bibfnamefont {R.~J.}\ \bibnamefont
  {Elliott}},\ }\bibfield  {title} {\bibinfo {title} {Intensity of optical
  absorption by excitons},\ }\href {https://doi.org/10.1103/PhysRev.108.1384}
  {\bibfield  {journal} {\bibinfo  {journal} {Phys. Rev.}\ }\textbf {\bibinfo
  {volume} {108}},\ \bibinfo {pages} {1384} (\bibinfo {year}
  {1957})}\BibitemShut {NoStop}%
\bibitem [{\citenamefont {Schöne}(2017)}]{Schone2017}%
  \BibitemOpen
  \bibfield  {author} {\bibinfo {author} {\bibfnamefont {F.}~\bibnamefont
  {Schöne}},\ }\emph {\bibinfo {title} {Optical Properties of yellow Excitons
  in Cuprous Oxide}},\ \href
  {https://link.aps.org/doi/10.1103/PhysRevLett.125.097401} {Ph.D. thesis},\
  \bibinfo  {school} {University of Rostock} (\bibinfo {year}
  {2017})\BibitemShut {NoStop}%
\bibitem [{\citenamefont {Fano}(1961)}]{Fano1961}%
  \BibitemOpen
  \bibfield  {author} {\bibinfo {author} {\bibfnamefont {U.}~\bibnamefont
  {Fano}},\ }\bibfield  {title} {\bibinfo {title} {Effects of configuration
  interaction on intensities and phase shifts},\ }\href
  {https://doi.org/10.1103/PhysRev.124.1866} {\bibfield  {journal} {\bibinfo
  {journal} {Phys. Rev.}\ }\textbf {\bibinfo {volume} {124}},\ \bibinfo {pages}
  {1866} (\bibinfo {year} {1961})}\BibitemShut {NoStop}%
\bibitem [{\citenamefont {Toyozawa}(1964)}]{Toyozawa1964}%
  \BibitemOpen
  \bibfield  {author} {\bibinfo {author} {\bibfnamefont {Y.}~\bibnamefont
  {Toyozawa}},\ }\bibfield  {title} {\bibinfo {title} {Interband effect of
  lattice vibrations in the exciton absorption spectra},\ }\href
  {https://doi.org/10.1016/0022-3697(64)90162-3} {\bibfield  {journal}
  {\bibinfo  {journal} {J. Phys. Chem. Solids}\ }\textbf {\bibinfo {volume}
  {25}},\ \bibinfo {pages} {59} (\bibinfo {year} {1964})}\BibitemShut {NoStop}%
\bibitem [{\citenamefont {Urbach}(1953)}]{Urbach1953}%
  \BibitemOpen
  \bibfield  {author} {\bibinfo {author} {\bibfnamefont {F.}~\bibnamefont
  {Urbach}},\ }\bibfield  {title} {\bibinfo {title} {The long-wavelength edge
  of photographic sensitivity and of the electronic absorption of solids},\
  }\href {https://doi.org/10.1103/PhysRev.92.1324} {\bibfield  {journal}
  {\bibinfo  {journal} {Phys. Rev.}\ }\textbf {\bibinfo {volume} {92}},\
  \bibinfo {pages} {1324} (\bibinfo {year} {1953})}\BibitemShut {NoStop}%
\bibitem [{\citenamefont {Cody}\ \emph {et~al.}(1981)\citenamefont {Cody},
  \citenamefont {Tiedje}, \citenamefont {Abeles}, \citenamefont {Brooks},\ and\
  \citenamefont {Goldstein}}]{Cody1981}%
  \BibitemOpen
  \bibfield  {author} {\bibinfo {author} {\bibfnamefont {G.~D.}\ \bibnamefont
  {Cody}}, \bibinfo {author} {\bibfnamefont {T.}~\bibnamefont {Tiedje}},
  \bibinfo {author} {\bibfnamefont {B.}~\bibnamefont {Abeles}}, \bibinfo
  {author} {\bibfnamefont {B.}~\bibnamefont {Brooks}},\ and\ \bibinfo {author}
  {\bibfnamefont {Y.}~\bibnamefont {Goldstein}},\ }\bibfield  {title} {\bibinfo
  {title} {Disorder and the optical-absorption edge of hydrogenated amorphous
  silicon},\ }\href {https://doi.org/10.1103/PhysRevLett.47.1480} {\bibfield
  {journal} {\bibinfo  {journal} {Phys. Rev. Lett.}\ }\textbf {\bibinfo
  {volume} {47}},\ \bibinfo {pages} {1480} (\bibinfo {year}
  {1981})}\BibitemShut {NoStop}%
\bibitem [{\citenamefont {Jang}\ \emph {et~al.}(2006)\citenamefont {Jang},
  \citenamefont {Sun}, \citenamefont {Watkins},\ and\ \citenamefont
  {Ketterson}}]{Jang2006}%
  \BibitemOpen
  \bibfield  {author} {\bibinfo {author} {\bibfnamefont {J.~I.}\ \bibnamefont
  {Jang}}, \bibinfo {author} {\bibfnamefont {Y.}~\bibnamefont {Sun}}, \bibinfo
  {author} {\bibfnamefont {B.}~\bibnamefont {Watkins}},\ and\ \bibinfo {author}
  {\bibfnamefont {J.~B.}\ \bibnamefont {Ketterson}},\ }\bibfield  {title}
  {\bibinfo {title} {Bound excitons in {Cu}${}_2${O}: Efficient internal free
  exciton detector},\ }\href {https://doi.org/10.1103/PhysRevB.74.235204}
  {\bibfield  {journal} {\bibinfo  {journal} {Phys. Rev. B}\ }\textbf {\bibinfo
  {volume} {74}},\ \bibinfo {pages} {235204} (\bibinfo {year}
  {2006})}\BibitemShut {NoStop}%
\bibitem [{\citenamefont {Schweiner}\ \emph {et~al.}(2017)\citenamefont
  {Schweiner}, \citenamefont {Main}, \citenamefont {Wunner},\ and\
  \citenamefont {Uihlein}}]{Schweiner2017}%
  \BibitemOpen
  \bibfield  {author} {\bibinfo {author} {\bibfnamefont {F.}~\bibnamefont
  {Schweiner}}, \bibinfo {author} {\bibfnamefont {J.}~\bibnamefont {Main}},
  \bibinfo {author} {\bibfnamefont {G.}~\bibnamefont {Wunner}},\ and\ \bibinfo
  {author} {\bibfnamefont {C.}~\bibnamefont {Uihlein}},\ }\bibfield  {title}
  {\bibinfo {title} {Even exciton series in {Cu}${}_2${O}},\ }\href
  {https://doi.org/10.1103/PhysRevB.95.195201} {\bibfield  {journal} {\bibinfo
  {journal} {Phys. Rev. B}\ }\textbf {\bibinfo {volume} {95}},\ \bibinfo
  {pages} {195201} (\bibinfo {year} {2017})}\BibitemShut {NoStop}%
\bibitem [{\citenamefont {Rommel}\ \emph {et~al.}(2021)\citenamefont {Rommel},
  \citenamefont {Main}, \citenamefont {Farenbruch}, \citenamefont {Yakovlev},\
  and\ \citenamefont {Bayer}}]{Rommel2021}%
  \BibitemOpen
  \bibfield  {author} {\bibinfo {author} {\bibfnamefont {P.}~\bibnamefont
  {Rommel}}, \bibinfo {author} {\bibfnamefont {J.}~\bibnamefont {Main}},
  \bibinfo {author} {\bibfnamefont {A.}~\bibnamefont {Farenbruch}}, \bibinfo
  {author} {\bibfnamefont {D.~R.}\ \bibnamefont {Yakovlev}},\ and\ \bibinfo
  {author} {\bibfnamefont {M.}~\bibnamefont {Bayer}},\ }\bibfield  {title}
  {\bibinfo {title} {Exchange interaction in the yellow exciton series of
  cuprous oxide},\ }\href {https://doi.org/10.1103/PhysRevB.103.075202}
  {\bibfield  {journal} {\bibinfo  {journal} {Phys. Rev. B}\ }\textbf {\bibinfo
  {volume} {103}},\ \bibinfo {pages} {075202} (\bibinfo {year}
  {2021})}\BibitemShut {NoStop}%
\bibitem [{\citenamefont {Konzelmann}\ \emph {et~al.}(2019)\citenamefont
  {Konzelmann}, \citenamefont {Frank},\ and\ \citenamefont
  {Giessen}}]{Konzelmann2019}%
  \BibitemOpen
  \bibfield  {author} {\bibinfo {author} {\bibfnamefont {A.}~\bibnamefont
  {Konzelmann}}, \bibinfo {author} {\bibfnamefont {B.}~\bibnamefont {Frank}},\
  and\ \bibinfo {author} {\bibfnamefont {H.}~\bibnamefont {Giessen}},\
  }\bibfield  {title} {\bibinfo {title} {Quantum confined rydberg excitons in
  reduced dimensions},\ }\href {https://doi.org/10.1088/1361-6455/ab56a9}
  {\bibfield  {journal} {\bibinfo  {journal} {Journal of Physics B: Atomic,
  Molecular and Optical Physics}\ }\textbf {\bibinfo {volume} {53}},\ \bibinfo
  {pages} {024001} (\bibinfo {year} {2019})}\BibitemShut {NoStop}%
\bibitem [{\citenamefont {Kavoulakis}\ \emph {et~al.}(1997)\citenamefont
  {Kavoulakis}, \citenamefont {Chang},\ and\ \citenamefont
  {Baym}}]{Kavoulakis1997}%
  \BibitemOpen
  \bibfield  {author} {\bibinfo {author} {\bibfnamefont {G.~M.}\ \bibnamefont
  {Kavoulakis}}, \bibinfo {author} {\bibfnamefont {Y.-C.}\ \bibnamefont
  {Chang}},\ and\ \bibinfo {author} {\bibfnamefont {G.}~\bibnamefont {Baym}},\
  }\bibfield  {title} {\bibinfo {title} {Fine structure of excitons in
  ${\mathrm{cu}}_{2}$o},\ }\href {https://doi.org/10.1103/PhysRevB.55.7593}
  {\bibfield  {journal} {\bibinfo  {journal} {Phys. Rev. B}\ }\textbf {\bibinfo
  {volume} {55}},\ \bibinfo {pages} {7593} (\bibinfo {year}
  {1997})}\BibitemShut {NoStop}%
\bibitem [{\citenamefont {Kr\"uger}\ \emph {et~al.}(2020)\citenamefont
  {Kr\"uger}, \citenamefont {Stolz},\ and\ \citenamefont
  {Scheel}}]{Kruger2020}%
  \BibitemOpen
  \bibfield  {author} {\bibinfo {author} {\bibfnamefont {S.~O.}\ \bibnamefont
  {Kr\"uger}}, \bibinfo {author} {\bibfnamefont {H.}~\bibnamefont {Stolz}},\
  and\ \bibinfo {author} {\bibfnamefont {S.}~\bibnamefont {Scheel}},\
  }\bibfield  {title} {\bibinfo {title} {Interaction of charged impurities and
  rydberg excitons in cuprous oxide},\ }\href
  {https://doi.org/10.1103/PhysRevB.101.235204} {\bibfield  {journal} {\bibinfo
   {journal} {Phys. Rev. B}\ }\textbf {\bibinfo {volume} {101}},\ \bibinfo
  {pages} {235204} (\bibinfo {year} {2020})}\BibitemShut {NoStop}%
\bibitem [{\citenamefont {Schlagm\"uller}\ \emph {et~al.}(2016)\citenamefont
  {Schlagm\"uller}, \citenamefont {Liebisch}, \citenamefont {Nguyen},
  \citenamefont {Lochead}, \citenamefont {Engel}, \citenamefont {B\"ottcher},
  \citenamefont {Westphal}, \citenamefont {Kleinbach}, \citenamefont {L\"ow},
  \citenamefont {Hofferberth}, \citenamefont {Pfau}, \citenamefont
  {P\'erez-R\'{\i}os},\ and\ \citenamefont {Greene}}]{Schlagmuller2016}%
  \BibitemOpen
  \bibfield  {author} {\bibinfo {author} {\bibfnamefont {M.}~\bibnamefont
  {Schlagm\"uller}}, \bibinfo {author} {\bibfnamefont {T.~C.}\ \bibnamefont
  {Liebisch}}, \bibinfo {author} {\bibfnamefont {H.}~\bibnamefont {Nguyen}},
  \bibinfo {author} {\bibfnamefont {G.}~\bibnamefont {Lochead}}, \bibinfo
  {author} {\bibfnamefont {F.}~\bibnamefont {Engel}}, \bibinfo {author}
  {\bibfnamefont {F.}~\bibnamefont {B\"ottcher}}, \bibinfo {author}
  {\bibfnamefont {K.~M.}\ \bibnamefont {Westphal}}, \bibinfo {author}
  {\bibfnamefont {K.~S.}\ \bibnamefont {Kleinbach}}, \bibinfo {author}
  {\bibfnamefont {R.}~\bibnamefont {L\"ow}}, \bibinfo {author} {\bibfnamefont
  {S.}~\bibnamefont {Hofferberth}}, \bibinfo {author} {\bibfnamefont
  {T.}~\bibnamefont {Pfau}}, \bibinfo {author} {\bibfnamefont {J.}~\bibnamefont
  {P\'erez-R\'{\i}os}},\ and\ \bibinfo {author} {\bibfnamefont {C.~H.}\
  \bibnamefont {Greene}},\ }\bibfield  {title} {\bibinfo {title} {Probing an
  electron scattering resonance using rydberg molecules within a dense and
  ultracold gas},\ }\href {https://doi.org/10.1103/PhysRevLett.116.053001}
  {\bibfield  {journal} {\bibinfo  {journal} {Phys. Rev. Lett.}\ }\textbf
  {\bibinfo {volume} {116}},\ \bibinfo {pages} {053001} (\bibinfo {year}
  {2016})}\BibitemShut {NoStop}%
\bibitem [{\citenamefont {Liebisch}\ \emph {et~al.}(2016)\citenamefont
  {Liebisch}, \citenamefont {Schlagmüller}, \citenamefont {Engel},
  \citenamefont {Nguyen}, \citenamefont {Balewski}, \citenamefont {Lochead},
  \citenamefont {Böttcher}, \citenamefont {Westphal}, \citenamefont
  {Kleinbach}, \citenamefont {Schmid}, \citenamefont {Gaj}, \citenamefont
  {Löw}, \citenamefont {Hofferberth}, \citenamefont {Pfau}, \citenamefont
  {P{\'{e}}rez-R{\'{\i}}os},\ and\ \citenamefont {Greene}}]{Liebisch2016}%
  \BibitemOpen
  \bibfield  {author} {\bibinfo {author} {\bibfnamefont {T.~C.}\ \bibnamefont
  {Liebisch}}, \bibinfo {author} {\bibfnamefont {M.}~\bibnamefont
  {Schlagmüller}}, \bibinfo {author} {\bibfnamefont {F.}~\bibnamefont
  {Engel}}, \bibinfo {author} {\bibfnamefont {H.}~\bibnamefont {Nguyen}},
  \bibinfo {author} {\bibfnamefont {J.}~\bibnamefont {Balewski}}, \bibinfo
  {author} {\bibfnamefont {G.}~\bibnamefont {Lochead}}, \bibinfo {author}
  {\bibfnamefont {F.}~\bibnamefont {Böttcher}}, \bibinfo {author}
  {\bibfnamefont {K.~M.}\ \bibnamefont {Westphal}}, \bibinfo {author}
  {\bibfnamefont {K.~S.}\ \bibnamefont {Kleinbach}}, \bibinfo {author}
  {\bibfnamefont {T.}~\bibnamefont {Schmid}}, \bibinfo {author} {\bibfnamefont
  {A.}~\bibnamefont {Gaj}}, \bibinfo {author} {\bibfnamefont {R.}~\bibnamefont
  {Löw}}, \bibinfo {author} {\bibfnamefont {S.}~\bibnamefont {Hofferberth}},
  \bibinfo {author} {\bibfnamefont {T.}~\bibnamefont {Pfau}}, \bibinfo {author}
  {\bibfnamefont {J.}~\bibnamefont {P{\'{e}}rez-R{\'{\i}}os}},\ and\ \bibinfo
  {author} {\bibfnamefont {C.~H.}\ \bibnamefont {Greene}},\ }\bibfield  {title}
  {\bibinfo {title} {Controlling rydberg atom excitations in dense background
  gases},\ }\href {https://doi.org/10.1088/0953-4075/49/18/182001} {\bibfield
  {journal} {\bibinfo  {journal} {Journal of Physics B: Atomic, Molecular and
  Optical Physics}\ }\textbf {\bibinfo {volume} {49}},\ \bibinfo {pages}
  {182001} (\bibinfo {year} {2016})}\BibitemShut {NoStop}%
\bibitem [{\citenamefont {Walther}\ and\ \citenamefont
  {Pohl}(2020)}]{Walther2020}%
  \BibitemOpen
  \bibfield  {author} {\bibinfo {author} {\bibfnamefont {V.}~\bibnamefont
  {Walther}}\ and\ \bibinfo {author} {\bibfnamefont {T.}~\bibnamefont {Pohl}},\
  }\bibfield  {title} {\bibinfo {title} {Plasma-enhanced interaction and
  optical nonlinearities of {Cu}${}_2${O} rydberg excitons},\ }\href
  {https://doi.org/10.1103/PhysRevLett.125.097401} {\bibfield  {journal}
  {\bibinfo  {journal} {Phys. Rev. Lett.}\ }\textbf {\bibinfo {volume} {125}},\
  \bibinfo {pages} {097401} (\bibinfo {year} {2020})}\BibitemShut {NoStop}%
\bibitem [{\citenamefont {Kitamura}\ \emph {et~al.}(2017)\citenamefont
  {Kitamura}, \citenamefont {Takahata},\ and\ \citenamefont
  {Naka}}]{Kitamura2017}%
  \BibitemOpen
  \bibfield  {author} {\bibinfo {author} {\bibfnamefont {T.}~\bibnamefont
  {Kitamura}}, \bibinfo {author} {\bibfnamefont {M.}~\bibnamefont {Takahata}},\
  and\ \bibinfo {author} {\bibfnamefont {N.}~\bibnamefont {Naka}},\ }\bibfield
  {title} {\bibinfo {title} {Quantum number dependence of the photoluminescence
  broadening of excitonic rydberg states in cuprous oxide},\ }\href
  {https://doi.org/10.1016/j.jlumin.2017.07.060} {\bibfield  {journal}
  {\bibinfo  {journal} {Journal of Luminescence}\ }\textbf {\bibinfo {volume}
  {192}},\ \bibinfo {pages} {808} (\bibinfo {year} {2017})}\BibitemShut
  {NoStop}%
\bibitem [{\citenamefont {Heck{\"o}tter}\ \emph {et~al.}(2021)\citenamefont
  {Heck{\"o}tter}, \citenamefont {Walther}, \citenamefont {Scheel},
  \citenamefont {Bayer}, \citenamefont {Pohl},\ and\ \citenamefont
  {A{\ss}mann}}]{Heckotter2021}%
  \BibitemOpen
  \bibfield  {author} {\bibinfo {author} {\bibfnamefont {J.}~\bibnamefont
  {Heck{\"o}tter}}, \bibinfo {author} {\bibfnamefont {V.}~\bibnamefont
  {Walther}}, \bibinfo {author} {\bibfnamefont {S.}~\bibnamefont {Scheel}},
  \bibinfo {author} {\bibfnamefont {M.}~\bibnamefont {Bayer}}, \bibinfo
  {author} {\bibfnamefont {T.}~\bibnamefont {Pohl}},\ and\ \bibinfo {author}
  {\bibfnamefont {M.}~\bibnamefont {A{\ss}mann}},\ }\bibfield  {title}
  {\bibinfo {title} {Asymmetric rydberg blockade of giant excitons in cuprous
  oxide},\ }\href {https://doi.org/10.1038/s41467-021-23852-z} {\bibfield
  {journal} {\bibinfo  {journal} {Nature Communications}\ }\textbf {\bibinfo
  {volume} {12}},\ \bibinfo {pages} {3556} (\bibinfo {year}
  {2021})}\BibitemShut {NoStop}%
\bibitem [{\citenamefont {Tokuda}\ \emph {et~al.}(2016)\citenamefont {Tokuda},
  \citenamefont {Nagata},\ and\ \citenamefont
  {Okada}}]{Tokuda2016SimultaneousEO}%
  \BibitemOpen
  \bibfield  {author} {\bibinfo {author} {\bibfnamefont {S.}~\bibnamefont
  {Tokuda}}, \bibinfo {author} {\bibfnamefont {K.}~\bibnamefont {Nagata}},\
  and\ \bibinfo {author} {\bibfnamefont {M.}~\bibnamefont {Okada}},\ }\bibfield
   {title} {\bibinfo {title} {Simultaneous estimation of noise variance and
  number of peaks in bayesian spectral deconvolution},\ }\href@noop {}
  {\bibfield  {journal} {\bibinfo  {journal} {ArXiv}\ }\textbf {\bibinfo
  {volume} {abs/1607.07590}} (\bibinfo {year} {2016})}\BibitemShut {NoStop}%
\bibitem [{\citenamefont {Iwamitsu}\ \emph {et~al.}(2016)\citenamefont
  {Iwamitsu}, \citenamefont {Aihara}, \citenamefont {Okada},\ and\
  \citenamefont {Akai}}]{Iwamitsu2016}%
  \BibitemOpen
  \bibfield  {author} {\bibinfo {author} {\bibfnamefont {K.}~\bibnamefont
  {Iwamitsu}}, \bibinfo {author} {\bibfnamefont {S.}~\bibnamefont {Aihara}},
  \bibinfo {author} {\bibfnamefont {M.}~\bibnamefont {Okada}},\ and\ \bibinfo
  {author} {\bibfnamefont {I.}~\bibnamefont {Akai}},\ }\bibfield  {title}
  {\bibinfo {title} {Bayesian analysis of an excitonic absorption spectrum in a
  {Cu}${}_2${O} thin film sandwiched by paired mgo plates},\ }\href
  {https://doi.org/10.7566/JPSJ.85.094716} {\bibfield  {journal} {\bibinfo
  {journal} {Journal of the Physical Society of Japan}\ }\textbf {\bibinfo
  {volume} {85}},\ \bibinfo {pages} {094716} (\bibinfo {year} {2016})},\
  \Eprint {https://arxiv.org/abs/https://doi.org/10.7566/JPSJ.85.094716}
  {https://doi.org/10.7566/JPSJ.85.094716} \BibitemShut {NoStop}%
\bibitem [{\citenamefont {Iwamitsu}\ \emph {et~al.}(2020)\citenamefont
  {Iwamitsu}, \citenamefont {Okada},\ and\ \citenamefont {Akai}}]{okada2020}%
  \BibitemOpen
  \bibfield  {author} {\bibinfo {author} {\bibfnamefont {K.}~\bibnamefont
  {Iwamitsu}}, \bibinfo {author} {\bibfnamefont {M.}~\bibnamefont {Okada}},\
  and\ \bibinfo {author} {\bibfnamefont {I.}~\bibnamefont {Akai}},\ }\bibfield
  {title} {\bibinfo {title} {Spectral decomposition of components weaker than
  noise intensity by bayesian spectroscopy},\ }\href
  {https://doi.org/10.7566/JPSJ.89.104004} {\bibfield  {journal} {\bibinfo
  {journal} {Journal of the Physical Society of Japan}\ }\textbf {\bibinfo
  {volume} {89}},\ \bibinfo {pages} {104004} (\bibinfo {year} {2020})},\
  \Eprint {https://arxiv.org/abs/https://doi.org/10.7566/JPSJ.89.104004}
  {https://doi.org/10.7566/JPSJ.89.104004} \BibitemShut {NoStop}%
\end{thebibliography}%

\end{document}